\documentclass[12pt,letterpaper,oneside, openany]{report}
\usepackage{amsfonts} 
\usepackage{amssymb}  
\usepackage{graphicx} 
\usepackage{amsmath}  
\usepackage[latin1]{inputenc}
\usepackage{latexsym}
\usepackage{color}
\usepackage{fancyhdr}
\usepackage[T1]{fontenc}
\usepackage[colorlinks=true]{hyperref}
\usepackage[all]{hypcap}
\usepackage{url}

\renewcommand{\tablename}{Tabla}

\newcommand{\fig}[1]{Fig~(\ref{#1})}

\begin{document}
\renewcommand{\tablename}{Tabla}

\pagenumbering{roman} \setcounter{page}{1}
\pagestyle{empty}

\begin{titlepage}
\begin{center}
{\large \bf UNIVERSIDAD DE LA HABANA.}
\end{center}

\begin{center}
\vspace{0.3cm}

{\large \bf FACULTAD DE FÍSICA.}

\vspace{0.3cm}

\begin{center}
\includegraphics[width=0.7\textwidth]{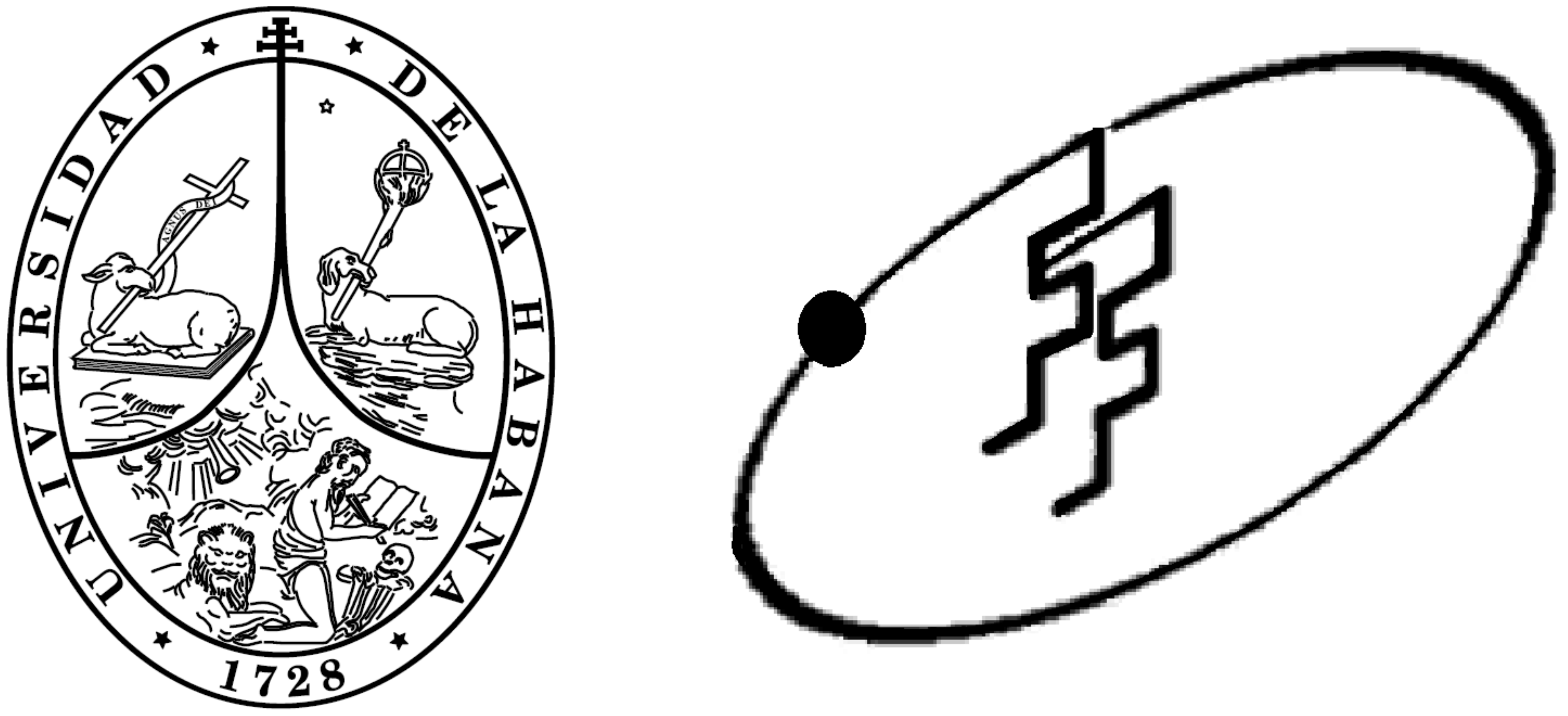}
\end{center}

\vspace*{0.3cm}

\begin{center}
{\Huge \bf Modelando las mutaciones en bacterias y tejidos humanos}
\end{center}

\vspace*{1.0cm}

{\large \bf Tesis de Diploma}\\
{\large  \it presentada en opci\'on al grado de} \\
{\Large  \bf Licenciado en F\'{\i}sica}\\
\end{center}
\vspace*{1.5cm}

{\large \bf Autor:}\hspace*{0.3cm}{\large Dario Alejandro León Valido, Facultad de Física.}

\vspace*{0.3cm}{\large \bf Tutor:}\hspace*{0.3cm}{\large Augusto González García, ICIMAF.}

\vspace*{1.0cm}

\begin{center}
\Large{La Habana, Cuba.}\\
\Large{2016}
\end{center}

\end{titlepage}


\newpage{}\voffset=2in \thispagestyle{empty} \hoffset=3.5in \textit{\Huge \ \ \ }\textit{\small {familia y amigos: lo logr\'e!!!}}{\small \par}

\vspace{10.5in}

\newpage
\hoffset=0in \voffset=1in \vspace{3in} \thispagestyle{empty} 
{\large\bf Agradecimientos}

En primer lugar debo agradecer al Dr. Rolando P\'erez y al Lic. Jorge Fern\'andez de Coss\'io, ambos pertenecientes al CIM, por acceder a formar parte 
del tribunal y realizar una muy buena oponencia en tiempo récord. Hicieron una exhaustiva revisión de la tesis, as\'i como de todos los art\'iculos  
citados; la cual conllevó un debate conmigo en  el CIM. Gracias por invitarme a almorzar, los que me conocen bien saben cuanto significa eso para m\'i. 

En especial, debo agradecer a mi tutor: Augusto Gonz\'alez, el cual no s\'olo ha sido mi modelo de f\'isico a seguir, sino también como persona. Me 
ha mostrado c\'omo  podemos interesarnos, opinar y actuar sobre diversas ramas de la sociedad, como el arte, la pol\'itica y la econom\'ia a un nivel 
m\'as riguroso y responsable para lograr aportar e influenciar en la toma de decisiones importantes para el desarrollo del pa\'is, sobretodo de la ciencia, 
rengl\'on altamente limitado en Cuba. Sigo las ideas planteadas en su bloc (http://augustoesm.blogspot.com/) y espero alcanzar suficiente madurez para que 
mis opiniones sean escuchadas y  respetadas tanto como las suyas.  

Le agradezco infinitamente a mi familia por apoyarme durante toda mi vida en cada cosa que se me ha ocurrido emprender. A mi madre, a mi hermana Daniela y 
a mi novia Emely por estar ah\'i conmigo en la realizaci\'on de la tesis, perdiendo horas de sueño, tecleando si hac\'ia falta y aliment\'andome (tarea 
sumamente dif\'icil).

A los profesores que me formaron, desde Etien en el 12 grado en la universidad con su rigurosidad extrema. Gretel, a quien le pregunt\'e sin cesar, casi 
todas las dudas que me surgieron en la carrera y, despu\'es de ese constante bombardeo, por ser parte del tribunal y no devolverme todo el fuego que me 
merec\'ia. A L\'idice, que adem\'as de ser una excelente profesora tambi\'en ha sido una muy buena amiga, s\'e que de haber sido falta hubiera ejercido 
como oponente de mi tesis. A todos los profesores de la carrera, siempre estar\'e dispuesto a devolver a las nuevas generaciones, todo lo que me han enseñado.

Al colectivo del ICIMAF, por permitir integrarme a su entorno de trabajo, un ambiente realmente agradable, en el cual desarroll\'e junto a mi tutor 
el grueso de la investigaci\'on. Por estar pendiente de mi bienestar e invitarme a cuanto evento se desarrollase.

A mis compañeros, por permanecer a mi lado durante toda la carrera. Realmente me divert\'i y la pas\'e bien y espero que podamos seguir haci\'endolo en lo 
adelante. Fue realmente refrescante nuestro estado de cuasi-felicidad, el cual nos permiti\'o hacer ``sugerencias'' durante las clases, jugar cancha despu\'es 
de haber salido m\'as o menos, bien o mal en las pruebas; Ale y Omar. Formar Incertidumbre Combinada, nuestro grupo de la facultad y seguir tocando a\'un 
cuando no sepamos c\'omo continuar; Alfredo. De alg\'un modo, seguir\'e haciendo esas cosas que tanto me gustan a la par de la F\'isica, ya que sin alguna 
de ellas no pudiera continuar.\newpage

Algunos no pudieron presenciar mi defensa, pero s\'e que estuvieron apoy\'andome desde la distancia. Mi padre, por ejemplo y mi buen amigo Heiner, quien 
desde el 12 grado me tra\'ia problemas interesantes para discutir conmigo, como  los Problemas del Milenio: 7 problemas matem\'aticos que su resoluci\'on 
ser\'ia premiada con 1 mill\'on de d\'olares cada uno. Dejando de lado el trofeo, su motivaci\'on ha sido impulsora de mi desarrollo intelectual.

Al resto de mis amigos, como Andr\'es, quien no s\'olo me proporcion\'o un algoritmo de ordenamiento de listas recurrente mucho m\'as eficiente del que yo 
me sab\'ia, sino que lo he molestado en su casa recurrentemente a cualquier hora. En fin, a todo aquel que de alguna manera ha estado pendiente de mi 
desarrollo y/o me ha brindado su ayuda, aunque no lo haya mencionado explícitamente o crea que su aporte es insignificante, si\'entase incluido cuando 
digo que esta tesis no hubiera sido posible sin USTEDES.  

\vspace{0.3in}


\newpage
\hoffset=0in \voffset=1in \vspace{3in} \thispagestyle{empty} {\large
\bf    Resumen}

Esta tesis tiene el propósito de estudiar las mutaciones, entendidas como trayectorias en el espacio de 
configuraciones del ADN (todas las combinaciones de bases). En ella se propone un modelo evolutivo para las 
mismas mediante trayectorias de Levy. Los par\'ametros del modelo se estiman a partir de datos provenientes 
de Experimentos de Evolución a Largo Plazo (EELP) con bacterias {\it E. Coli.} A partir de dicho modelo se simulan varios resultados
de los experimentos como la aparición de nuevos genotipos y su competencia. La selecci\'on natural se incluye en el modelo 
a través del parámetro de {\it fitness}, el cual caracteriza la ``salud'' de cada genotipo bajo determinadas condiciones ambientales. 
Adem\'as, se discute cualitativamente la analog\'ia 
encontrada entre los fenotipos mutantes de las bacterias y las c\'elulas cancerosas. Se analiza el papel de la radiaci\'on como fuente
de mutaciones, en especial la proveniente de la desintegraci\'on del rad\'on presente en el aire que respiramos.

\vspace{0.3in}

\hoffset=0in \voffset=1in \vspace{0.5in} \thispagestyle{empty}
{\large \bf Abstract}

This thesis is aimed at studying mutations, understood as trajectories in the DNA configuration space.
An evolutive model of mutations in terms of Levy flights is proposed. The parameters of the model are estimated
by means of data from the Long-Term Evolution Experiment (LTEE) with {\it E. Coli} bacteria. The results of simulations on
competition of clones, mean fitness, etc are compared with experimental data. We discuss the qualitative analogy found between
the bacterial mutator phenotype and the cancerous cells. The role of radiation as source of mutations is analyzed. We focus on the case
of Radon's decay in the lungs in breathing. 

\vspace{0.3in}



\voffset=0in 




\tableofcontents
\pagestyle{myheadings}
\setcounter{page}{0} \pagebreak \pagenumbering{arabic}
\setcounter{page}{1}
\pagestyle{myheadings}
\chapter*{Pr\'ologo}\label{cap0}
\addcontentsline{toc}{chapter}{Pr\'ologo}

Esta tesis contiene nuestras ideas de c\'omo concebir las mutaciones, que son el fruto de m\'as de 2 a\~nos de trabajo, estudio
y b\'usqueda de informaci\'on. A lo largo de este proceso, hemos tenido la necesidad de estudiar temas de biolog\'ia, para entender la enorme cantidad
de procesos biol\'ogicos que subyacen de trasfondo en nuestro problema. La mayor\'ia de ellos son explicados en el primer cap\'itulo introductorio de manera
simple y enfocados en los principales aspectos que nos interesa desarrollar, por lo que pudiera servir de gu\'ia para introducirse en esta extensa e interesante 
rama de la ciencia, que es la 
Biolog\'ia. Varias secciones de la tesis est\'an pensadas con el objetivo de que sirvan de base para el desarrollo de futuras investigaciones, ya que a\'un 
hay mucho por hacer y descubrir en lo referente a las mutaciones. Como podr\'a apreciar el lector en el segundo cap\'itulo, esta tesis pudiera ser ejemplo, en mi
m\'as humilde opini\'on, de c\'omo nosotros, f\'isicos al fin, podemos investigar temas tan diversos y alejados de nuestra formaci\'on inicial. Tambi\'en,
de nuestra necesidad de pasar de descripciones cualitativas a modelos matem\'aticos, para entender mejor la naturaleza de los sistemas.


\chapter{Introducci\'on}\label{cap1}

\section{Mutaciones en bacterias}\label{cap11}

\subsection{El ciclo del desarrollo celular. Replicación del ADN}\label{cap111}

Las células se reproducen duplicando su contenido y luego dividiéndose en dos. 
En especies unicelulares como las bacterias, cada división produce un organismo adicional. Los detalles del 
ciclo en cada caso pueden variar, pero los requerimientos son universales. En primer lugar, se 
debe replicar satisfactoriamente el ADN de la célula progenitora y luego los cromosomas replicados deben 
segregarse en células separadas. El ciclo de división involucra un conjunto de procesos, los cuales son agrupados por fases. 
La vasta mayoría de las células también duplica sus masas y sus orgánulos, para ello deben estar coordinados una serie de procesos 
citoplasmáticos y nucleares. Existe un sistema de control del ciclo que regula el proceso como un todo. 
Nuestro interés está encaminado a entender cómo se producen los errores genéticos que pasan a otras generaciones, 
por tanto, debemos enfocarnos en el proceso de replicación del ADN nuclear que ocurre durante una parte 
específica de la interfase: la fase S, y en el papel que juega el sistema de control. 

El sistema de control del ciclo celular es un dispositivo bioquímico que opera cíclicamente. 
Está construido a partir de un grupo de proteínas interactuantes que inducen y coordinan los procesos de forma jerárquica. 
Dichas proteínas pueden detener al ciclo en puntos de chequeo específicos, reunir información para verificar el 
cumplimiento de la jerarquía, retroalimentarse y actuar en consecuencia; todo mediante mecanismos bioquímicos. 
Las pausas también son importantes para permitir cierta regulación del sistema de control. 
Esta regulación se da generalmente en dos puntos: justo antes de iniciar la fase S y a la entrada de la mitosis, 
donde se realiza el reparto equitativo del material hereditario \cite{BioMol}.

\begin{figure}
\begin{center}
\includegraphics[width=13cm]{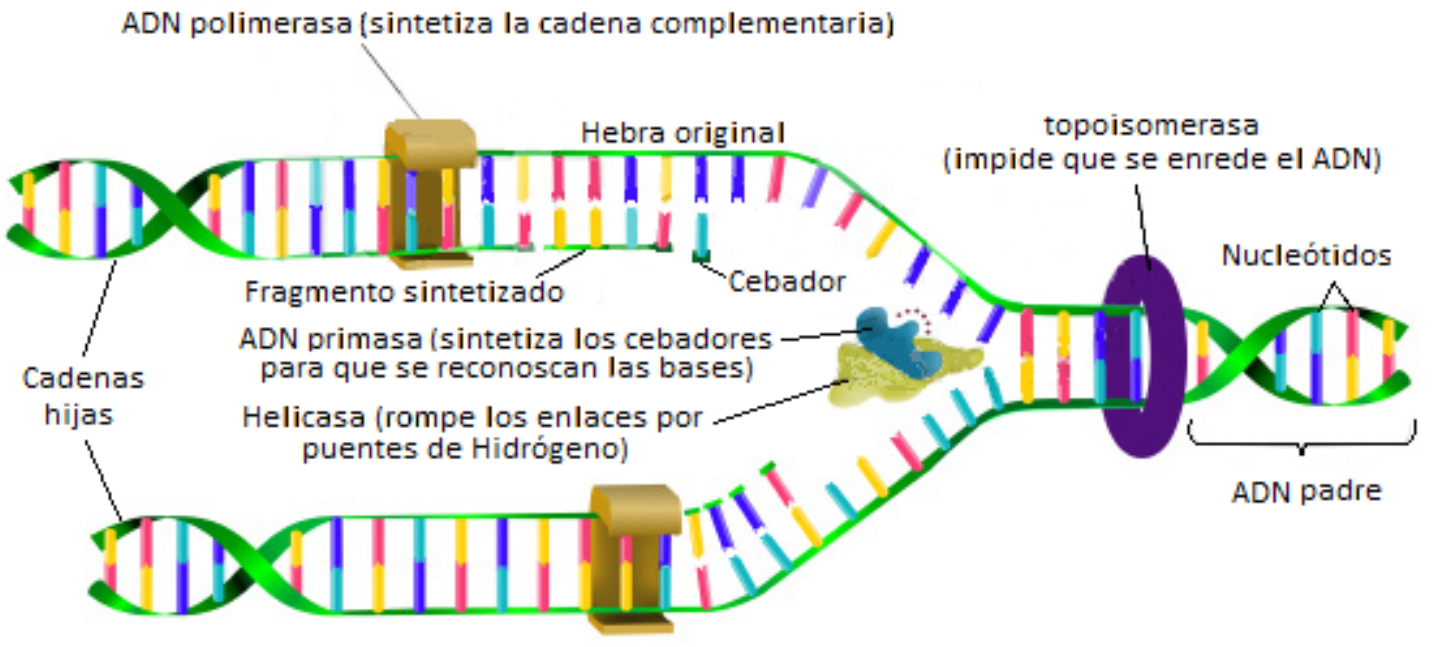}
\caption{Algunos procesos y enzimas que participan en la replicaci\'on del ADN. } \label{Replication}
\end{center}
\end{figure}

La replicación del ADN ocurre a ritmos de polimerización cercanos a 500 nucleótidos por segundo en bacterias y a 
50 en mamíferos. Claramente las proteínas que la catalizan son rápidas y precisas. Estas propiedades se logran mediante 
complejas multienzimas que guían el proceso y constituyen una elaborada ``máquina de replicación''. Primero se desenrolla 
la estructura de doble hélice que forman las dos cadenas de bases del ADN. Luego se eliminan los enlaces por puentes de 
hidrógeno existentes entre las bases complementarias de cada cadena. Una vez separadas, cada cadena hace de plantilla 
para la síntesis de una nueva (\fig{Replication}). Como cada molécula de ADN hija se queda con una cadena de la madre, el proceso de replicación 
se dice que es semiconservativo. Gracias a la complementación existente entre las bases que forman al ADN, las moléculas 
hijas serán iguales a la madre.  Durante la replicación actúan moléculas que evitan que el ADN se enrede.

\subsection{Mecanismos de corrección y mutaciones}\label{cap112}

Como todo proceso celular, la replicación se basa en el reconocimiento molecular \cite{BioMol}, cuyo primer paso es la difusión. 
Los encuentros entre moléculas ocurren aleatoriamente por simple difusión térmica, es por eso que los procesos de 
reconocimiento no son perfectos. La estructura de doble hélice del ADN es un mecanismo efectivo para garantizar una 
buena copia, pero no está libre de errores: ocurren alrededor de uno por $10^4$ o $10^5$ bases \cite{BioMol}. Este valor es asombrosamente 
pequeño, no obstante, esta tasa de error no pudiera mantener la integridad del genoma humano que es de 
aproximadamente $3\times10^9$ bases \cite{BioMol}. Existen mecanismos de reparación del ADN encargados de mejorar la fidelidad de la copia. 
Por ejemplo, luego de la duplicación del ADN, bloques de re-replicación aseguran que ningún segmento sea copiado más de 
una vez. Estos bloques son retirados al pasar a la mitosis. Existe otro mecanismo de reparación de fallas que difiere de 
la mayoría de los sistemas de reparación del ADN, el cual no depende de la presencia de nucleótidos anormales que puedan 
ser detectados y extirpados. En cambio, detecta distorsiones en el exterior de la hélice que resultan de un apareamiento 
de bases no compaginado. El resultado final de la copia con los arreglos es de un error en $10^9$ bases adicionadas \cite{BioMol}. Muy 
raramente la máquina de replicación se salta o adiciona nucleótidos, o pone T por C, o A por G. Cada tipo de 
cambio del ADN constituye una mutación que se copia en las siguientes generaciones como la secuencia correcta. A los 
intercambios puntuales de bases que emergen de la replicación se les llama mutaciones ``espontáneas'' y tienen un carácter local.

\subsection{Mutaciones y Selección Natural}\label{cap113}

En un proceso evolutivo celular no siempre podremos notar todas las mutaciones que se han producido en instantes anteriores, 
ya que una mutación en un gen puede inactivar una proteína crucial y causar la muerte de la célula, entonces se pierde. Las 
mutaciones que dis\-mi\-nu\-yen grandemente la capacidad del individuo para sobrevivir y/o reproducirse son llamadas ``deletéreas''. 
Por otro lado, puede ocurrir una mutación ``silenciosa'' fuera del sitio activo de la proteína y no afectar su función 
correspondiente, permaneciendo inactiva. Rara vez una mutación crea un gen con una mejora o función útil. En este caso 
los organismos mutantes tienen una ventaja y el gen mutado eventualmente reemplaza al original por selección natural. 
Estas mutaciones se fijan y pueden ser observadas en la gran mayoría de la descendencia \cite{BioMol, 20mil_Lenski, Adaptation_Lenski}, 
constituyen una nueva característica 
del genotipo de la especie. La influencia del ambiente es esencial en el proceso selectivo porque es quien dicta las reglas 
de la competencia, determina significativamente las características fenotípicas de los organismos y también, de manera 
indirecta a través de las mutaciones, las características genotípicas promedio de la población. Existe un parámetro 
llamado  {\it fitness} que describe la supervivencia de cada 
genotipo emergente de la evolución. Un genotipo con más ``salud'' aumentará su frecuencia de aparición en las generaciones 
siguientes, por tanto, el {\it fitness} es un parámetro que caracteriza la evolución de las mutaciones.

En la literatura existen numerosas definiciones de {\it fitness}, sin embargo, todas ellas concuerdan aproximadamente en la 
esencia de la idea \cite{Fitness}: el {\it fitness} involucra la habilidad de los organismos o, más raramente, poblaciones o especies 
para sobrevivir y reproducirse en el medio ambiente en el cual están inmersos \cite{Fitness}. Por tanto, podemos referirnos al 
{\it fitness} individual, colectivo de una población de organismos o relativo entre individuos o poblaciones.

\subsection{Factores externos en las mutaciones}\label{cap114}

Existen mutaciones no espontáneas que no provienen de errores en el proceso de replicación. Ellas involucran mutágenos: 
agentes químicos, biológicos o físicos que alteran la información genética de un organismo, lo cual incrementa la frecuencia 
de mutaciones por encima del nivel natural \cite{BioMol}. Generalmente producen cambios no locales que involucran un reajuste (inserción, 
perdida, inversión, etc.) de un segmento de la molécula de ADN \cite{BioMol}. Los mutágenos biológicos son aquellos organismos vivos que 
pueden alterar la secuencia del material genético de su hospedador como los virus. Los mutágenos químicos 
son compuestos capaces de alterar las estructuras del ADN de forma brusca. Los ejemplos más cotidianos de mutágenos físicos 
son las distintas radiaciones que reciben los organismos y que pueden romper los enlaces de la molécula de ADN. En condiciones 
normales los mutágenos químicos y biológicos son mucho más controlables; mientras que los físicos, aun aislando a los organismos 
de la radiación externa, están presentes en la desintegración de algunas de las partículas componentes de los mismos.

\subsection{El experimento de evolución a largo plazo en poblaciones de {\it \textbf{E. Coli}}}\label{cap115}

El profesor R. Lenski y su grupo, en la Universidad de Michigan han llevado a cabo un experimento de evolución a largo plazo con 
bacterias {\it E. Coli} por casi 30 
años, con el objetivo de estudiar las dinámicas fenotípicas y genotípicas de poblaciones de bacterias. Cada día las bacterias 
se replican 6 o 7 generaciones de evolución binaria, alcanzando 3 400 generaciones en un año, lo que representa poblaciones 
crecidas por más de 60 000 generaciones hasta la fecha \cite{Sitio_Lenski}. En el experimento, 12 cultivos de bacterias con un ancestro común 
evolucionan independientemente. Cada día 0.1 ml del caldo bacterial es transferido rigurosamente a 9.9 ml de solución de glucosa, 
mantenido a temperatura controlada hasta el día siguiente. En las primeras 8 horas el número de bacterias varía aproximadamente 
de acuerdo con la ley $N_0 2^t/t_0$, que es consistente con la forma de reproducción: la división celular por mitosis. Luego que la glucosa se consuma,
se alcanza un estado estacionario: en las restantes 16h no hay mortalidad apreciable \cite{Sitio_Lenski}. 

Entre los resultados del experimento se encuentran el {\it fitness} relativo a la cepa ancestral \cite{Adaptation_Lenski}, datos sobre
mutaciones puntuales \cite{A1,A2} y sobre reajustes grandes en los cromosomas \cite{mBio}. Más adelante estos datos serán utilizados en el trabajo.

\section{Mutaciones en tejidos humanos}

\subsection{Identidad del ADN del individuo}\label{cap121}

El ADN del genoma de un organismo, que es el mismo en todas sus c\'elulas, puede dividirse conceptualmente en dos: el que codifica proteínas 
(los genes) y el que no.
En la mayoría de las especies pluricelulares, al contrario de las bacterias, sólo una pequeña fracción del genoma 
codifica proteínas. Por ejemplo, alrededor del $1.5 \%$ 
del genoma humano consiste en exones que codifican proteínas (20 000 a 25 000 genes), mientras que más del $90\%$ consiste en 
ADN no codificante \cite{BioMol}. Existen regiones no codificantes que aparentemente no tienen función alguna: el llamado ``ADN basura''.
Otras secuencias no codificantes tienen afinidad hacia proteínas especiales que tienen la capacidad de unirse al ADN, con un 
papel importante en el control de los mecanismos de trascripción y replicación. A estas secuencias se les llama frecuentemente 
``secuencias reguladoras'', porque están encargadas de la activación o la supresión de la expresión genética. Ellas
no codifican proteínas, pero sí se transcriben a ARN: ribosómico, de transferencia y de interferencia (ARNi, que son los
ARN que bloquean la expresión de genes específicos) \cite{BioMol}.
Por otro lado, algunas secuencias de ADN desempeñan un papel estructural en los cromosomas: los telómeros y centrómeros contienen 
pocos o ningún gen codificante de proteínas, pero son importantes para estabilizar la estructura de los cromosomas.  Las estructuras
de intrones y exones de algunos genes, son importantes por 
permitir los cortes y empalmes alternativos del pre-ARN mensajero que hacen posible la síntesis de diferentes proteínas a partir 
de un mismo gen (sin esta capacidad no existiría el sistema inmune, por ejemplo) \cite{BioMol}. Algunas secuencias de ADN no codificante 
representan pseudogenes que tienen valor evolutivo, ya que permiten la creación de nuevos genes con nuevas funciones. Otros ADN no 
codificantes proceden de la duplicación de pequeñas regiones del ADN \cite{BioMol, Sitio_Lenski}, lo que pudiera ser otro mecanismo contra la aparición de 
mutaciones, ya que los daños quedarían repartidos entre m\'as bases y las regiones afectadas pueden ser reparadas a partir de sus
duplicados. El rastreo de secuencias repetidas en el ADN permite estudios de relaciones de parentesco (filogenética) \cite{BioMol}.

\subsection{División celular en organismos pluricelulares. Especialización}\label{cap122}

Los organismos pluricelulares pueden ser vistos como una sociedad o ecosistema donde cada miembro individual es una célula, los 
cuales se reproducen por división celular y se organizan en ensambles de colaboración o tejidos \cite{BioMol}. Dado la gran cantidad de células 
que poseen estos organismos, hacen falta muchas rondas de divisiones para que, a partir de una, surja un nuevo individuo. Muchos 
organismos como aves y mamíferos dejan de crecer cuando alcanzan un tamaño determinado, pero, incluso cuando el crecimiento se 
detiene, siguen proliferando células \cite{BioMol}. En la adultez, la división celular también es necesaria para reemplazar células desgastadas 
por su constante uso, perdidas por heridas o que sufrieron una muerte programada (apoptosis), de modo que el organismo como un todo 
permanece invariante \cite{BioMol}. La apoptosis puede darse como mecanismo de eliminación de células inservibles para el tejido porque están dañadas 
y no cumplen con su función, para autorregular su crecimiento y desarrollo o, simplemente, para controlar la cantidad de células en el 
tejido \cite{BioMol}. Un humano adulto puede producir millones de células nuevas en un segundo para mantener su estado de equilibrio y, si se detuvieran 
todas sus divisiones celulares, moriría en unos días \cite{BioMol}. En un instante dado la mayoría de las células en la adultez no están creciendo o 
dividiéndose, en cambio están en un estado de descanso: perfeccionando su función especializada mientras se retiran del ciclo celular. \cite{BioMol}

La mayoría de las poblaciones de células diferenciadas en vertebrados están muriendo y siendo reemplazadas continuamente. La regeneración 
requiere de un crecimiento coordinado de los componentes del tejido. Las nuevas células pueden ser producidas de dos maneras: por 
replicación simple de células del mismo tipo (sin cambios de especialización) o, por células madres relativamente no diferenciadas 
en procesos que involucran cambios del fenotipo de las células \cite{BioMol}. La diferenciación se va dando de manera gradual en la descendencia 
a partir de cambios epigenéticos: que no modifican al ADN; no por mutaciones, que también están presentes \cite{BioMol}. Este proceso es posible 
gracias a que los sectores no codificantes del ADN regulan la activación de los genes y, estos pueden ser silenciados a conveniencia para 
que la célula cumpla una función determinada \cite{BioMol}. El principal mecanismo de la epigenética es la transferencia de grupos metilos a algunas 
de las bases (metilación) para alterar la transcripción genética \cite{BioMol}. 

Se requieren células madres dondequiera que exista una necesidad recurrente de reemplazar células diferenciadas que no puedan dividirse 
por sí mismas. Este es el caso de la incompatibilidad del estado final de la diferenciación con la división celular en muchos tejidos. La 
poca diferenciabilidad de las células madres radica en que no se encuentran al final de una cadena de especialización \cite{BioMol}. Las células madres 
también pueden dividirse sin límites hasta el final de la vida del organismo. Cada una de sus hijas puede permanecer como célula madre o 
embarcarse en un curso irreversible hacia el final de la diferenciación \cite{BioMol}. Las tasas de renovación varían de un tejido a otro: pueden ser 
de días, como las paredes epiteliales del intestino delgado (renovadas por células madres), o de un año y más, como en el páncreas 
(renovadas por simple duplicación) \cite{BioMol, Tomasetti}. Muchos tejidos con bajas tasas de renovación pueden ser estimulados para producir nuevas células en 
grandes cantidades cuando haga falta, por ejemplo en casos de infección \cite{BioMol}.

No todas las poblaciones de células especializadas en el cuerpo están sujetas a renovación. Algunas son generadas en el embrión en números 
apropiados y se retienen hasta la adultez \cite{BioMol}. Al parecer, estas poblaciones nunca se dividen y no pueden ser reemplazadas si se pierden. En 
mamíferos son de este tipo casi todas las células nerviosas, las células musculares del corazón, las células de los pelos auditivos del 
oído y las células de la lente del ojo. Todas estas células tienen largos períodos de vida y necesariamente viven en ambientes protegidos. 
Son diferentes en otros sentidos y no se sabe una razón en general de por qué deberían ser irremplazables. Hay muy pocas tan inmutables 
como las fibras de las lentes, las demás son capaces de renovar sus partes componentes a pesar de no replicarse \cite{BioMol}. 

\subsection{Mecanismos de corrección a nivel celular y el sistema inmuno\-ló\-gi\-co} \label{cap123}

El daño en el ADN inicia una respuesta que activa diferentes mecanismos de reparación que reconocen lesiones específicas en la 
molécula y son reparadas en el momento \cite{BioMol}. Alternativamente, si el daño genómico es demasiado grande, la célula puede caer en un estado 
irreversible de inactividad llamado senescencia \cite{BioMol}. En casos extremos de daños no reparables, los me\-ca\-nis\-mos de control a nivel 
celular inducirán la activación de una serie de rutas celulares que culminarán en la muerte celular \cite{BioMol}. Según el tipo de daños que 
ha sufrido la estructura de doble hélice del ADN, han evolucionado una variedad de estrategias de reparación que restauran la 
información perdida \cite{BioMol}. Si es posible, las células utilizan la cadena de ADN complementaria como ``plantilla'' para restaurar la 
información original de manera fidedigna \cite{BioMol}. Las roturas de cadena doble son especialmente peligrosas para la célula, ya que pueden 
provocar mutaciones en el genoma y son más difíciles de corregir. Existen dos mecanismos que reparan estas roturas: la unión de 
extremos no homólogos y la reparación recombinativa (también conocida como reparación asistida por plantilla o reparación de 
recombinación homóloga) \cite{BioMol}. El primer mecanismo busca secuencias homólogas cortas presentes en las colas de los extremos de ADN 
que deben ser unidos. Si estas secuencias son compatibles, la reparación suele ser correcta, pero también puede causar mutaciones 
durante la misma. La reparación recombinante requiere la presencia de una secuencia idéntica o casi idéntica que sea utilizada 
como plantilla para reparar la rotura. La maquinaria enzimática responsable de este proceso es muy parecida a la del cruce cromosómico 
durante la meiosis en la reproducción sexual \cite{BioMol}. Esta ruta permite que un cromosoma dañado sea reparado utilizando una cromátida hermana 
o un cromosoma homólogo como plantilla. Algunas roturas de cadena doble pueden ser causadas por intentos de replicación en una molécula 
de ADN con una rotura de cadena única o una lesión no reparada. Estos casos son reparados generalmente por recombinación \cite{BioMol}. 

Otro posible destino de células con un elevado daño genómico es convertirse en cancerosas. Estas células tienen un grado de mutabilidad 
que le permite hacer cambios en su metabolismo y su proliferación de manera autónoma, burlando los controles de regulación del 
metabolismo y de la división celular del tejido \cite{BioMol}. Existen genes en el ADN que se utilizan para eliminarlas, los llamados genes 
supresores de tumores \cite{BioMol}. A este nivel, las células cancerosas son tratadas como cualquier otro agente patógeno introducido en el 
organismo, que produce daños y enfermedades. Para combatir estos males los vertebrados tienen una maquinaria de defensa muy bien 
desarrollada llamada sistema inmunitario \cite{BioMol}. El sistema inmune está compuesto por muchos tipos de proteínas, células, órganos y 
tejidos, los cuales se relacionan en una red elaborada y dinámica. Dicho sistema protege a los organismos de las infecciones con 
varias líneas de defensa \cite{BioMol}. Las más simples son las barreras físicas, que evitan que patógenos como bacterias y virus entren en el 
organismo. Si un patógeno penetra estas barreras, el sistema inmunitario innato ofrece una respuesta inmediata, pero no específica. 
Si los agentes patógenos evaden la respuesta innata, existe una tercera capa de protección: el sistema inmunitario adaptativo \cite{BioMol}. 
En él, se adapta la respuesta durante la infección para mejorar el reconocimiento del agente patógeno. Esta información se conserva 
aún después de que el agente sea eliminado, bajo la forma de memoria inmunitaria \cite{BioMol}. La capacidad de memorizar le permite al sistema 
inmune desencadenar ataques más rápidos y más fuertes en el futuro si se detectan los mismos tipos de patógenos. La inmunidad de 
un organismo depende de la habilidad de su sistema inmunitario para distinguir entre las moléculas propias y las que no lo son. 
Para ello se utiliza un tipo de células llamadas linfocitos, las cuales contienen moléculas receptoras que reconocen objetivos o 
blancos específicos \cite{BioMol}. En la respuesta inmediata se libera una clase de linfocitos llamados células B, las cuales producen anticuerpos 
estándares para la mayoría de las infecciones. En la inmunidad adaptativa se utilizan los linfocitos llamados células T, las cuales 
reconocen un objetivo no propio, sólo después de procesar pequeños fragmentos del patógeno (antígenos).  La respuesta se elabora 
combinando el antígeno con receptores propios que se obtienen de la transcripción de fragmentos de ADN seleccionados, en principio, 
aleatoriamente \cite{BioMol}.

\subsection{Cáncer}\label{cap124}

Anteriormente hemos descrito la aparición de células cancerosas a partir de mutaciones, pero, en nuestra visión social de las 
células de un organismo, hemos omitido un factor fundamental en su evolución: la selección natural. No mencionamos ninguna 
competencia entre las células resultantes del proceso de división. La razón es que, en este sentido, un cuerpo saludable es 
una sociedad muy particular en la que el auto-sacrificio es una regla mucho más fuerte que la competencia \cite{BioMol}. Todos los linajes 
de células somáticas están destinados a morir sin dejar descendencia; dedicando su existencia, solo a sostener la de las 
células germinales, las cuales por sí mismas no podrían sobrevivir. Por tanto, a diferencia de células como las bacterias, 
las células de un organismo están obligadas a colaborar. Cualquier mutación que de riesgos de comportamientos egoístas en los 
miembros de la cooperativa celular pondrá en peligro el futuro de toda la empresa.

El cáncer es una enfermedad en la cual células mutantes individuales (cancerosas) comienzan a prosperar a costa de sus vecinos, 
se reproducen descontroladamente y al final terminan destruyendo toda la sociedad celular; por tanto, sus ingredientes básicos 
son las mutaciones, la competencia y la selección natural \cite{BioMol}. Se clasifican de acuerdo al tejido u órgano del cual nacen. El desarrollo 
del cáncer como un proceso micro evolucionario ocurre en escalas temporales de meses o años \cite{BioMol}. Sus características difieren en 
dependencia del tipo de célula a partir de la cual se derivó. También tienen la propiedad hereditable de poder invadir y colonizar 
te\-rri\-to\-rios normalmente reservados para otras células (metástasis) \cite{BioMol}. Incluso cuando un cáncer ha metastizado, se puede seguir un 
rastro genético hasta su origen: generalmente un tumor singular primario que surgió en cierto tejido \cite{BioMol}. En la mayoría de estas pruebas 
se presume que el tumor primario se derivó de la división celular de una única célula, la cual sufrió algún cambio hereditario en 
su ADN que le permitió sobrepasar a sus vecinos. Evidentemente, una mutación puntual no es suficiente para convertir una célula 
típicamente saludable en cancerosa. Muchas líneas de evidencias indican que la génesis del cáncer requiere que ocurran varios 
accidentes raros en una misma célula \cite{BioMol}. Mucha evidencia proviene de estudios epidemiológicos de la incidencia de cáncer como 
función de la edad. Si una sola mutación fuera responsable, ocurriendo con una probabilidad fija por año, la oportunidad de desarrollar 
cáncer en cualquier año debería ser independiente de la edad. Para la mayoría de los cánceres es un hecho que la probabilidad
crece estrepitosamente con la edad, típicamente como una potencia cuadrada, cúbica, cuarta o quinta \cite{BioMol}. Para esas estadísticas
se estima que deben haber ocurrido entre 3 y 7 eventos aleatoriamente independientes, cada uno de baja probabilidad, que son 
típicamente requeridos para convertir una célula en cancerosa. Los números más pequeños se aplican a casos de leucemia (cáncer
en la sangre, no se forma tumor), mientras que los mayores se aplican a carcinomas (cáncer en células epiteliales o glandulares) \cite{BioMol}.

\subsection{Envejecimiento y cáncer}\label{cap125}

Teniendo en cuenta la influencia de factores externos en las mutaciones, podemos encontrar muchos cánceres cuyas ocurrencias 
correlacionan con una alta exposición de los individuos a alguno de estos factores. Para cánceres que poseen una causa externa 
apreciablemente discernible, casi siempre existe un tiempo de retraso grande entre los eventos causales y el comienzo de la 
enfermedad \cite{BioMol}. Por ejemplo, el cáncer de pulmón no comienza a crecer vertiginosamente hasta dentro de 10 o 20 años después de que 
la persona comenzara a fumar en gran medida, exponiéndose a sustancias carcinogénicas producidas en la combustión de los 
cigarrillos. También, la incidencia de leucemia en Hiroshima y Nagasaki no muestran un aumento considerable de casos hasta 5 
años después de la explosión de las bombas atómicas, el máximo índice de enfermos se da transcurridos 8 años \cite{BioMol}.  Pudiéramos citar 
más ejemplos. Durante este largo período de incubación las eventuales células cancerosas sufren una sucesión de cambios. La 
mayoría de las personas mueren de algún otro achaque antes de que el cáncer tenga tiempo de desarrollarse, pero no significa 
que no hayan cultivado células cancerosas durante su vida. Por el contrario, la frecuencia de grandes mutaciones y la tasa de 
eliminación de células dañadas en el organismo son bastante altas, lo que indica que constantemente se producen y eliminan un 
gran número de células cancerosas en el cuerpo \cite{BioMol}. Es una guerra sin fin entre el tumor queriendo crecer y el sistema inmune 
debilitándolo. Por eso, es necesario comprender cómo los distintos factores intervienen en la lucha, posibilitando o no, la 
aparición de la enfermedad. Un factor elemental es el envejecimiento, no solo porque al pasar más tiempo se da chance a que 
aparezcan más mutaciones, sino también porque nuestra salud empeora y nuestras defensas decaen. Por estas razones son más frecuentes 
los canceres en edades avanzadas.

\subsection{Datos sobre riesgo de cáncer en tejidos}\label{cap126}

\begin{figure}
\begin{center}
\includegraphics[width=13cm]{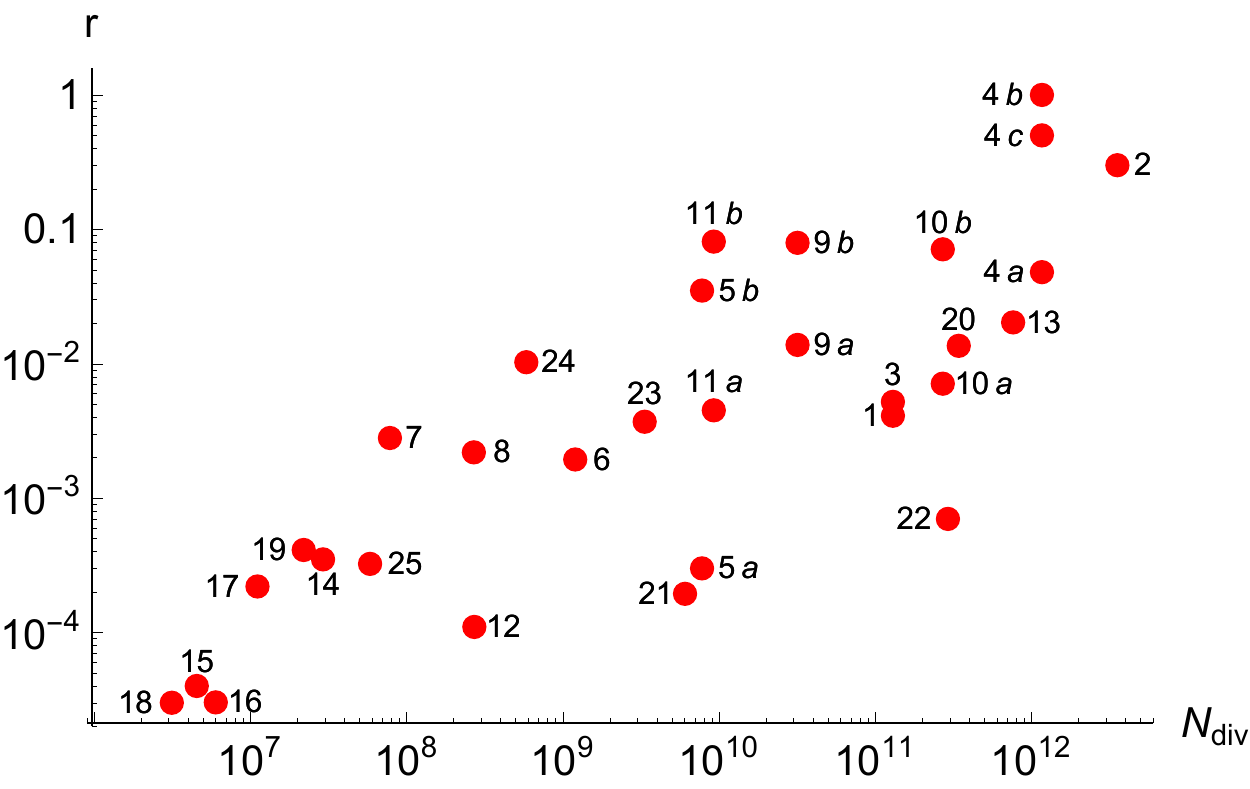}
\caption{Relaci\'on etre el n\'umero de divisiones de las c\'elulas madres en la vida de un tejido dado y su correspondiente 
riesgo de c\'ancer. 1: leucemia mieloide aguda, 2: carcinoma basocelular, 3: leucemia linf\'atica cr\'onica, 4a: c\'ancer colorrectal, 
4b: c\'ancer colorrectal con PAF, 4c: c\'ancer colorrectal con s\'indrome de Lynch, 5a: c\'ancer de duodeno, 5b: c\'ancer de duodeno 
con PAF,  6: c\'ancer de c\'elulas escamosas del es\'ofago, 7: c\'ancer de ves\'icula no papilario, 8: glioblastoma, 9a: carcinoma de c\'elulas 
escamosas de la cabeza y el cuello, 9b: carcinoma de c\'elulas escamosas de la cabeza y el cuello con VPH-16, 10a: hepatocarcinoma, 10b: 
hepatocarcinoma con hepatitis C, 11a: adenocarcinoma de pulm\'on (no fumadores), 11b: adenocarcinoma de pulm\'on (fumadores), 12: 
meduloblastoma, 13: melanoma, 14: osteosarcoma, 15: osteosarcoma de los brazos, 16: osteosarcoma de la cabeza, 17: osteosarcoma de las 
piernas, 18: osteosarcoma de la pelvis, 19: c\'elulas germinales del ovario, 20: adenocarcinoma de c\'elulas de los conductos pancre\'aticos, 
21: carcinoma neuroendocrino pancre\'atico, 22: adenocarcinoma del intestino delgado, 23: c\'elulas germinales de los test\'iculos, 24: 
carcinoma folicular/papilar de tiroides y 25: carcinoma de tiroides medular.
 } \label{plotTomas}
\end{center}
\end{figure}

Es bien conocido que existen grandes diferencias de incidencia de cáncer en los distintos tejidos del cuerpo. De los casi 
100 tipos de cáncer detectados en humanos, los más comunes en hombres son los de pulmón, próstata, colorrectal y del estómago; 
mientras que en mujeres son los de mama, colorrectal, de pulmón y de cervical. Si se incluyera el cáncer de piel, en total 
representarían cerca del $40\%$ de los casos \cite{BioMol}. Algunas de las diferencias están asociadas a factores de riesgo ya estudiados, 
como fumar, el uso abusivo del alcohol, radiación ultravioleta, o virus como el de papiloma humano. Pero esto se aplica solo 
a poblaciones expuestas a potentes mutágenos. Si comparamos datos de riesgo de padecer cáncer en la vida tomados de \cite{Tomasetti}, esas 
exposiciones no pudieran explicar las diferencias entre, por ejemplo, los valores correspondientes a los tejidos del tracto 
alimenticio. Existe un factor de 24 entre el intestino grueso y el delgado, cuando supuestamente están expuestos a condiciones 
similares. En dicho trabajo se propone que la razón de estas diferencias es la presencia de otro factor: el efecto estocástico 
asociado al número de replicaciones que hacen las células madres de cada tejido en la vida de un individuo. Como existen órdenes 
de diferencia entre las frecuencias de replicación de las células madres de cada tejido, el riesgo de cáncer variará a consecuencia. 
Para demostrar que los efectos estocásticos constituyen un factor de riesgo de cáncer importante y pueden ser separados de la 
herencia y los factores ambientales, ellos calculan la correlación entre el riesgo (medido estadísticamente en una población) y 
estimados numéricos de dichos efectos. Los datos son mostrados en la \fig{plotTomas}.

Si denotamos por r al riesgo y por $N_{div}$ al número de divisiones en un tejido, podemos estimar $N_{div}$ de la siguiente manera:

\begin{equation}
 N_{div}=N_{cel} f_d \tau_v
\end{equation}

\noindent
donde $N_{cel}$ es la cantidad de células madres presentes en el tejido (permanece constante debido al proceso de regulación de la 
mitosis), $f_d$ es la frecuencia de división celular de cada tejido (cantidad de divisiones por año) y $\tau_v$  es el tiempo de vida 
medio de una persona (80 años). 

Los datos de la \fig{plotTomas} fueron incluidos en el Anexo 1 junto a otros datos de inter\'es de los tejidos, que también fueron tomados de \cite{Tomasetti}.


\section{Radiaciones ionizantes y mutaciones}\label{cap2}


\subsection{Dosis de radiación}

La magnitud que caracteriza la incidencia de radiación sobre un cuerpo es la dosis de radiación. Existen tres tipos \cite{w}: la dosis absorbida $D$,
la cual se define como la energía absorbida por unidad de masa y en el Sistema Internacional se da en Gray (J/Kg); la dosis equivalente,
es la suma de las contribuciones a la dosis de todos los tipos de radiación absorbida teniendo en cuenta que algunas son más dañinas que
otras: 

\begin{equation}
H =\sum_{R}w_R D_R, 
\end{equation}

\noindent
 $w_R$ es un factor adimensional que caracteriza a cada radiación (alfa, beta, ...) y en el S.I. $H$ se da en 
Sievert (Sv). La otra dosis es la dosis efectiva que está referida a un tejido en específico:

\begin{equation}
E_T = w_T H_T, 
\end{equation}

\noindent
donde $H_T$ es la dosis equivalente que recibe el tejido y $w_T$ es un factor adimensional de peso que refleja la sensibilidad del tejido a la 
radiación. Está definido de tal forma que si $H$ se distribuye homogéneamente por los tejidos, la suma de las dosis 
efectivas de los tejidos será igual a la dosis equivalente:

\begin{equation}
\sum_{T}w_T = 1. 
\end{equation}

De las tres dosis existentes la única medible es la absorbida, porque los factores $w_R$ y $w_T$ son cualitativos y son estimados a partir de datos 
experimentales, sobre todo a partir de datos provenientes de desastres nucleares de gran escala, por lo que pueden sufrir modificaciones 
según el aumento de los datos experimentales, en especial $w_T$. En el Anexo 2 aparecen datos de estos coeficientes para algunos 
tejidos, as\'i como de los de la radiaci\'on, que fueron tomados de la Ref \cite{w}.

\subsection{Dosis en los diferentes tejidos}

Una vez conocidas las diferentes magnitudes que intervienen en la medición de los daños producidos por las diferentes radiaciones ionizantes sobre el cuerpo, 
debemos hacer un estudio detallado de su incidencia sobre los distintos tejidos humanos. De esta forma pudiéramos determinar cuáles son los más afectados, 
así como las fuentes más importantes. En la referencia \cite{Rep160} se hizo un minucioso estudio sobre este aspecto, al calcular estimados de dosis anuales que reciben 
todos los tejidos del cuerpo y analizar las fuentes más importantes de radiaciones. Los autores describen detalladamente todos sus procedimientos.  En la 
tabla \ref{tabla} se 
muestran datos de dosis anuales en dos años diferentes, que recibe nuestro organismo en nuestra vida cotidiana (tomados de la Ref. \cite{Rep160}). Las fuentes 
son clasificadas 
en naturales, donde se incluyen las principales radiaciones que nos llegan del medio ambiente; médicas, donde se toman en cuenta las distintas radiaciones 
que recibimos de dispositivos tecnológicos de la medicina que se utilizan para hacer diagnósticos o determinados tratamientos y el resto (consumo, industrial 
y de ocupación), son radiaciones provenientes de nuestra propia alimentación, del uso de algunas sustancias, de nuestro entorno de trabajo o de procesos 
industriales.  

Como es de esperar, en los dos años hemos recibido aproximadamente la misma dosis de radiación natural. Sin embargo, en los veinte años de diferencia entre 
las dos mediciones, la dosis médica ha aumentado vertiginosamente producto del gran desarrollo tecnológico alcanzado. La dosis que recibimos de fuentes no 
naturales prácticamente ha hecho duplicar la dosis total que recibimos anualmente. 

\begin{table}[!ht] \label{tabla}
\begin{center}
\begin{tabular}{|c|l c|l c|}\hline
\multicolumn{5}{|c|}{\textbf{Reporte No.160 del NCRP, datos de E.E.U.U.}}\\ \hline\hline
\textbf{Dosis} && \textbf{1984} && \textbf{2004}\\ \hline\hline
Natural& \textbf{3.0} &\hspace{1mm} 2.0  Rad\'on y Torio &\textbf{3.11}&2.28 Rad\'on y Torio\\
&& 1.0 Otros &&0.83 Otros\\ \hline
&& 0.39 Diagn\'osticos && 1.47 Tomograf\'ias\\ 
M\'edica&\textbf{0.53}&  &\textbf{3.0}& 0.77 Med. Nuclear\\ 
&&0.14 Med. Nuclear&&0.76 Otros\\ \hline
Consumida&&0.05-0.13&&0.13\\ \hline
Industrial&&0.001&&0.001\\ \hline
Ocupacional&&0.009&&0.005\\ \hline\hline
\textbf{Total} && \textbf{3.6 mSv/a\~no} && \textbf{6.2 mSv/a\~no} \\ \hline
\end{tabular}
\caption{Reporte No.160 del NCRP, datos de E.E.U.U de la dosis de radiaci\'on anual recibida por persona.}
\end{center}
\end{table}

\subsection{Datos sobre la concentraci\'on de Rad\'on y c\'ancer}

De los datos anteriores podemos llegar a la conclusión que el pulmón es el órgano que se lleva la mayor parte de la dosis de radiación que recibe el cuerpo 
humano. Pr\'acticamente todo el rad\'on y torio y una buena parte de la que viene de procedimientos m\'edicos. Con las radiaciones ionizantes que m\'as lidiamos 
en la cotidianidad, teniendo en cuenta su 
grado de malignidad, resultan de la desintegración del radón o sus hijos, también radiactivos.  Su isótopo más abundante es el $^{222}Rn$, con una abundancia 
natural del $100\%$. Este se origina a partir de reacciones nucleares en las capas m\'as altas de la atm\'osfera, contaminando el aire que respiramos. Además,
es un gas noble, por lo que es poco reactivo y no se pierde en ninguna reacci\'on atmosf\'erica. 
Otros afluentes de su radiactividad provienen de cadenas de desintegración de elementos más pesados presentes en la tierra, como el is\'otopo de uranio: $^{238}U$.
En el reporte de Ref \cite{Radon} también se informa la presencia de este is\'otopo en materiales constructivos, explicando así, la gran diferencia de actividad 
medida dentro y fuera de las viviendas (entiéndase cualquier construcción). Este elemento es gaseoso, pero la larga lista de sus productos radiactivos son 
metales (s\'olidos), por lo que la c\'elula impactada por la primera desintegración recibe un da\~no adicional considerable.

\begin{figure}
\begin{center}
\includegraphics[width=11cm]{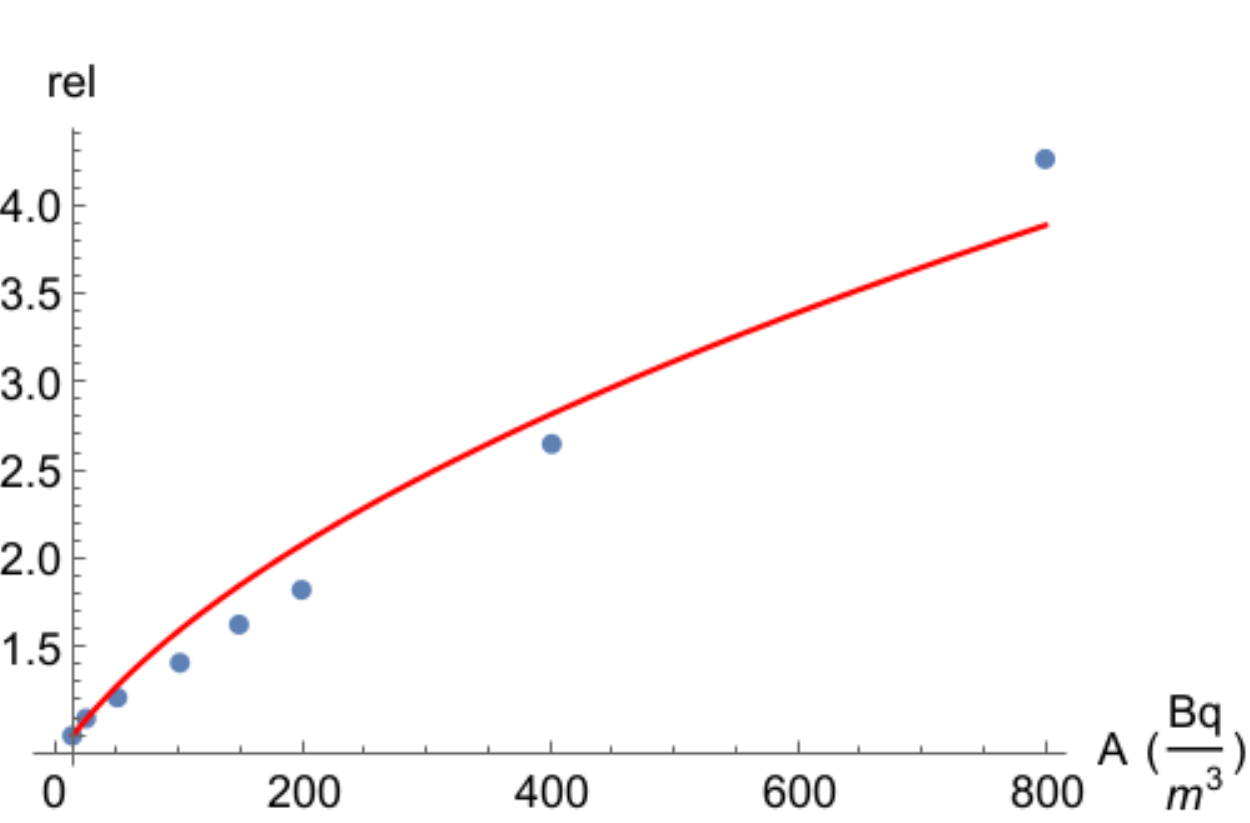}
\caption{ Aumento del riesgo de c\'ancer en funci\'on de la concentraci\'on de rad\'on a la que estamos expuestos relativo al riesgo cuando no hay exposici\'on.
La l\'inea continua representa un ajuste de los datos. } \label{riesgoR}
\end{center}
\end{figure}

Dada la gran influencia que ejerce el radón en nuestro cuerpo y bajo sospechas de posible causa de cáncer, los autores de la Ref \cite{Radon} realizaron un 
minucioso estudio en pos de relacionar las altas concentraciones de radón en la atm\'osfera con la ocurrencia de cáncer. Ellos midieron su concentración 
en distintas ciudades de los E.E.U.U. con la intención de correlacionarla con el índice de casos de cáncer de pulmón (riesgo de padecer la enfermedad en la vida) 
en dichas localidades. Sus resultados son mostrados en la \fig{riesgoR}. Resulta curioso el alto grado de similitud que presentan con los datos de riesgo de cáncer en 
tejidos en función del número de divisiones de células madres de cada uno, presentados en \fig{plotTomas}. Esto indica que la radiación en general es otro
factor de riesgo, aunque aparentemente de menor peso, que deber\'ia tenerse en cuenta para mejorar la correlaci\'on de los datos de la \fig{plotTomas}.

En la \fig{riesgoR} también se muestra un ajuste de los datos a partir de una dependencia potencial, que se obtiene de un modelo desarrollado para las mutaciones 
en el Cap\'itulo 2. También, pudieran ser ajustados a una recta para calcular la probabilidad de ocurrencia de c\'ancer
debido al radón. Algo parecido se hace con los datos de la \fig{plotTomas} en la sección \ref{Xx}.


\chapter{Mutaciones y trayectorias de Levy}\label{Cap3}

\section{El carácter acumulativo de las mutaciones}

En el Capítulo \ref{cap1} nos referimos a la baja frecuencia ($10^{-9}$) con que se producen mutaciones puntuales en el ADN, como resultado de fallas
en la etapa de replicación. Algunos cambios en el ADN pueden ser producidos en otras etapas del ciclo celular debido a procesos intr\'insecos; 
como radicales libres dentro de la c\'elula, o agentes externos, como radiaciones y sustancias t\'oxicas. Si estos cambios pasan la replicación
del ADN y se heredan, también dan lugar a mutaciones.

En bacterias y otros organismos simples, las mutaciones son ``filtradas'' por el mecanismo de selección natural. S\'olo los fenotipos m\'as 
aptos persisten. En el hombre, por el contrario, la mayor\'ia de las c\'elulas tienen una vida muy corta y solamente los linajes de c\'elulas
madres pueden acumular cambios en el ADN, a lo largo de la vida del individuo.

\begin{figure}
\begin{center}
\includegraphics[width=13cm]{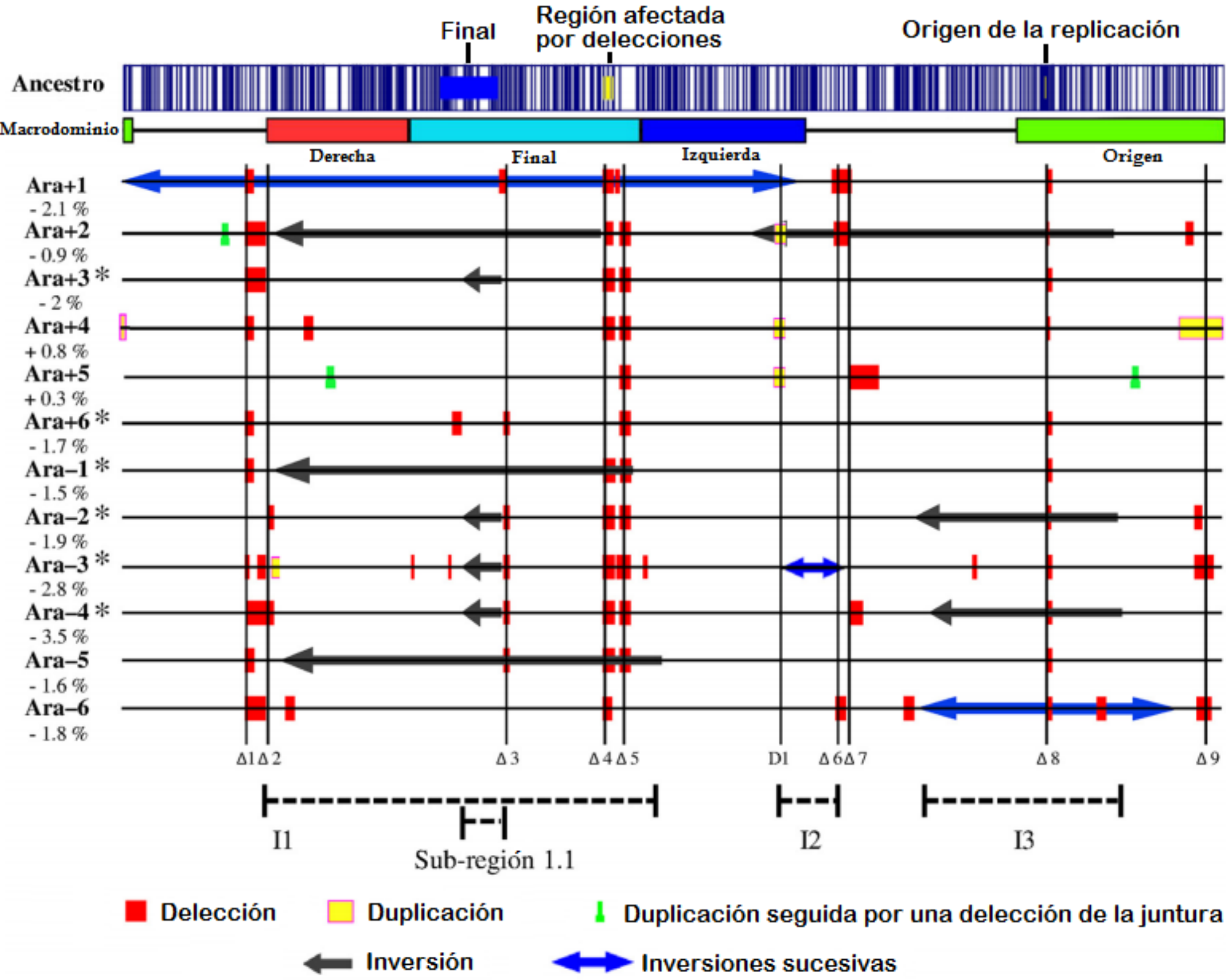}
\caption{\cite{mBio} Reajustes de gran escala en cromosomas de clones de muestras de cada una de las 12 poblaciones de {\it E. Coli} (Ara-1-Ara-6, Ara+1-Ara+6) del 
EELP, que evolucionaron por 40 000 generaciones. El asterisco marca los clones que evolucionaron con tasas de mutación superiores a las vistas en su 
antepasado. El valor de porciento mostrado representa el cambio relativo del tamaño del clon respecto a su ancestro. El mapa \'optico de
la cadena ancestral es mostrado en la parte superior, en donde se señalan las localizaciones del origen y el final de la replicación, junto
con una zona que fue afectada por delecciones en 10 de los clones. Los dominios cromosomales se indican debajo del mapa. Todos los reajustes
se muestran de manera relativa al genoma ancestral usando el c\'odigo de colores mostrado debajo de la figura. Las l\'ineas verticales 
nombradas $\Delta 1$-$\Delta 9$ indican regiones afectadas repetitivamente (en dos o m\'as poblaciones) por delecciones, mientras que 
la l\'inea $D1$ indica una región afectada repetitivamente por eventos de duplicaci\'on. Tres intervalos cromosómicos, mostrados como 
I1-I3, y la subregi\'on (1.1) sin I1 fueron afectados repetidamente por inversiones.} \label{MapaMut}
\end{center}
\end{figure}

En el proceso acumulativo,
las mutaciones que involucran reajustes a gran escala pudieran ser fundamentales. Los tipos de reajustes más comunes son perdidas de 
fragmentos (delecciones), inserciones, desplazamientos (translocaciones) e inversiones. Estas últimas consisten en intercambiar las hebras 
de la doble cadena de ADN del fragmento y, aunque es un cambio de varias bases, no se pierde información, pero produce problemas en la codificación. 
Podemos analizar, por ejemplo, la ocurrencia de grandes mutaciones en las 12 poblaciones de bacterias del EELP \cite{mBio} como se muestra en la \fig{MapaMut}. 

Los autores definen un reajuste en paralelo en poblaciones, como aquellos que involucran regiones cromosómicas que fueron afectadas por el mismo 
tipo de evento de reajuste: delecciones, inversiones, duplicaciones, etc. en al menos dos poblaciones distintas. El análisis de reajustes 
en paralelo en clones de poblaciones donde aparecen fenotipos mutantes, pudiera resolver qué mutaciones especificas le hacen falta al mutante. Sin 
embargo, los reajustes que se encuentran en común están presentes en casi todas las poblaciones, por lo que no lo caracterizan. La mayoría 
son delecciones de fragmentos duplicados, pedazos previamente introducidos por algún virus, otros sin ninguna función conocida (``basura'') o alguna mutación 
ligeramente beneficiosa \cite{mBio}.

La presencia de reajustes en paralelo en las poblaciones del EELP que evolucionan independientemente, constituye una evidencia muy fuerte del carácter 
acumulativo de las mutaciones y, de cierta manera, sugiere tener en cuenta el número total de mutaciones en cada generación.

\section{Las mutaciones como trayectorias en el espacio de las configuraciones del ADN}

Considerando una variable $x$ que caracterice todas las posibles modificaciones del ADN y teniendo en cuenta la acumulabilidad de las mutaciones, 
podemos analizar su evolución temporal en las células. En el proceso de replicación, representando el parentesco de las células por líneas 
genealógicas, se determinan trayectorias desde la o las, células iniciales hasta las que sobreviven al final. Estas trayectorias son análogas 
a las definidas en \cite{Long_Tail} y \cite{Levy_Cancer}. La idea de trayectorias en un proceso evolutivo celular significa que existen cadenas de 
Markov de mutaciones \cite{Probability}, donde los cambios en el ADN de una célula, medidos por $x$, en el paso $i+1$ están dados por los cambios en 
el paso anterior $i$ más una modificación adicional $\delta$:

\begin{equation}
x_{i+1}=x_i+\delta 
\end{equation}

\noindent
El parámetro $\delta$ representa las mutaciones en el paso $i+1$ y no es el cambio producido por factores endógenos o externos, sino el cambio resultante 
después de actuar los mecanismos de reparación. 

\section{Midiendo los cambios en el ADN}

Es difícil englobar con una variable todos los tipos de variaciones que ocurren en el ADN de manera efectiva. Si pensamos en las mutaciones puntuales, 
una forma muy conveniente de medir dichos cambios es definiendo una caminata sobre el ADN, la cual se emplea en \cite{Landscape} para describir el propio paisaje 
fractal de la molécula. Este método también pudiera ser usado para el resto de las mutaciones. Podemos utilizar una variante similar a la de ellos. 
Primero, definimos una variable auxiliar en el sitio $\alpha$ de la molécula: $u_\alpha(A)=1/8$, $u_\alpha(G)=3/8$, $u_\alpha(C)=-3/8$ y $u_\alpha(T)=-1/8$. 
Luego, una caminata a lo largo del ADN queda definida de la siguiente manera: 

\begin{equation}
 y(l)=\sum_{\alpha = 1}^{l}u_\alpha
\end{equation}

Como función de $l$, la variable $y$ describe un perfil de la molécula de ADN. En \cite{Walks} ca\-rac\-te\-ri\-zan este perfil a trav\'es de las
fluctuaciones medias, descritas cuantitativamente por la desviación cuadr\'atica media: $F(l)=\sqrt{\langle\Delta y^2(l)\rangle - {\langle\Delta y(l)\rangle}^2}$, la cual está
dada en función de la cantidad $\Delta y(l)$, definida como $\Delta y(l)=y(l+l_0)-y(l_0)$. La media se calcula por todas las posibles longitudes $l_0$. 
Ellos encontraron una diferencia sustancial de comportamiento entre la región del ADN que codifica y la que no, en genes de c\'elulas eucariotas. 
En la zona codificante encontraron que los nucleótidos están poco correlacionados entre sí. Este resultado es típico de procesos no correlacionados o 
de correlación de corto alcance, donde se puede definir una longitud característica de correlación, como se espera de la distribución normal del 
movimiento browniano. La función $F(l)$, la cual define el radio browniano, sigue una ley $l^{1/2}$ en estos casos \cite{Walks}. A veces, la zona no codificante, 
en contraste, exhibe correlaciones de largo alcance (libre de escala) cuya ley, $l^a$, presenta coeficientes distintos: $a \ne 0.5$. 

A diferencia de \cite{Walks}, donde s\'olo se estudiaba la forma del ADN mediante la variable $y(l)$, nosotros tambi\'en estamos interesados en su evoluci\'on 
temporal, acorde con lo expuesto en la sección anterior. Sus modificaciones pueden ser medidas como: $y(l)-y_0(l)$, donde $y$ corresponde al ADN mutado 
y $y_0$ a la configuración inicial. Para caracterizar al gran número de componentes de $y$ ($5\times 10^9$ bases) la estrategia puede ser utilizar 
variables que midan los cambios globales o distancias a la función original:

\begin{equation}
 x(L)=\sum_{\alpha = 1}^{L}(u'_\alpha-u_\alpha),
\end{equation}

\begin{equation}
 x^{(1)}(L)=\sum_{\alpha = 1}^{L}(u'_\alpha-u_\alpha),
\end{equation}

\noindent
$x^{(2)}$ (el segundo momento), la entrop\'ia de Shannon \cite{Entropy}, etc. $L$ es la longitud de la molécula. 

Tomando la primera variante por ejemplo, una sustitución de citosina (C) por guanina (G) en alg\'un punto se corresponder\'a con un valor $x(L) = 3/4$, 
mientras que todos los reemplazamientos puntuales producen cambios en promedio de 5/12. En casos de intercambios de varias bases, existen reconfiguraciones que producen 
los mismos valores de $x(L)$, pero podemos diferenciarlos llevando el n\'umero de bases cambiadas como variable adicional. A\'un as\'i, son evidentes los
grandes problemas que presentan estas propuestas para describir todos los fen\'omenos de las mutaciones.

La razón de que exista una diferencia de comportamiento entre las dos zonas del ADN, está dada por el hecho de que mutaciones 
más dañinas en la zona de codificación afectan la transcripción de proteínas vitales, por tanto, o destruyen la célula o se corrige el error, en 
cualquier caso no se observan con la misma frecuencia. 

La caminata aleatoria tiene serios problemas para medir los cambios de más de una base en el ADN, por ejemplo; si se eliminan, se insertan o se trasladan 
segmentos con bases al azar, el simple hecho de introducir cambios no modifica la correlación. S\'olo en caso de inserciones de segmentos con bases 
correlacionadas la cambiaría, las delecciones y las translocaciones nunca lo harían. Sin embargo, de los resultados de \cite{Walks} se pueden inferir varias 
cosas. Primero, que las mutaciones no se manifiestan de la misma forma en las zonas codificantes y no codificantes, si no tendrían perfiles semejantes. 
Segundo, dado que la caminata aleatoria resuelve bien las mutaciones puntuales, entonces el movimiento browniano debe ser una buena descripción de las 
mismas, porque reproduce su carácter local. Tercero, debe existir otra componente $\delta_{SL}$, que caracterice a las mutaciones de mayor tamaño, que no se rigen por el movimiento 
browniano, y debe ser libre de escala. La combinación del movimiento browniano y los saltos de largo alcance es característica de los vuelos de Levy. 
Por lo que, a pesar de los inconvenientes que aparecen al intentar medir los cambios en el ADN, pensamos que la evolución de las mutaciones debe regirse 
por un proceso de Levy. De ellos hablaremos más adelante.

Los resultados de \cite{Walks} también motivaron al desarrollado de un modelo de vuelos de Levy generalizados para la secuencia de bases del ADN 
propuesto en \cite{Levy_Walks}.

\section{Trayectorias de Levy}

Un vuelo de Levy es un tipo de paseo aleatorio en el cual los incrementos son distribuidos de acuerdo a una distribución de probabilidad de cola pesada. 
Específicamente, la distribución usada es una ley potencial de la forma $y = x^{-a}$ donde $1 < a < 3$ \cite{Levy_Flights}. El escalamiento en forma de ley de 
potencias de las longitudes de pasos, da a los vuelos de Levy una propiedad de escala invariante, es decir, tienen una varianza infinita. Esta 
característica es propia de un fractal \cite{Levy_Flights}.

Luego de discutir temas referentes a la medici\'on de los cambios en el ADN en la sección anterior, podemos considerar, por cuesti\'on de 
simplicidad, una \'unica variable unidimensional $x_L(t)$, donde $L$ es el tamaño de la cadena de ADN. El tiempo, por otro lado, se representa mejor
por el n\'umero de generaciones a lo largo de un linaje celular, medido desde el ancestro común. En una población en evolución, el patr\'on dibujado por 
el conjunto de variables $\{x_L(t)\}$, una para cada c\'elula, resulta la distribución de probabilidad. Para generar la distribución en nuestro modelo, 
partimos de una cadena de Markov con el siguiente incremento:

\begin{equation}
 \delta = \delta_B + \delta_{SL},
\end{equation}

\noindent
donde $\delta_B$ es la componente browniana. 
La otra componente no local se caracteriza por eventos raros de probabilidad total $p<<1$ y densidad de probabilidad proporcional a $1/\delta_{SL}^\nu$, 
con $1<\nu<3$. M\'as adelante estimaremos el exponente.
Notemos que para la descripción de los saltos largos se requieren variables adicionales, por ejemplo el largo $L$ de la cadena, el cual cambia después de una 
inserci\'on o una delecci\'on de un fragmento de ADN. Para la probabilidad de saltos largos usamos el ansatz $p=p_{SL}\pi(l)$, donde $p_{SL}$ es el ritmo con
que se producen (eventos/generaciones) y $\pi(l)$ es la probabilidad normalidada de un evento que involucra un segmento de longitud $l \geq 1$. Al separar
la parte temporal de la probabilidad de saltos largos, asumimos que los eventos se encuentran uniformemente distribuidos en el tiempo. De acuerdo a
los vuelos de Levy: $\pi(l)=(\nu-1)/l^{\nu}$, donde $1<\nu<3$. 

Pudi\'eramos preguntarnos por qu\'e los vuelos de Levy deber\'ian describir las mutaciones. Puede haber un argumento general en favor de este modelo.
En el EELP, donde el tamaño de la poblaci\'on es controlado, la evolución biol\'ogica puede verse como un problema de optimizaci\'on. El {\it fitness}
medio de la población es la función a maximizar. Las mutaciones proveen un mecanismo para buscar extremos en el espacio de las configuraciones, mientras que 
la selección natural escoge las mejores representaciones de la población. Una b\'usqueda local, como las mutaciones puntuales, pudieran atrapar las
trayectorias de mutaciones alrededor de un m\'aximo local, visto en el paisaje del {\it fitness} como función de la configuración. Un algoritmo de b\'usqueda 
m\'as optimizado deber\'a incluir saltos largos de cualquier tamaño, como una distribución libre de escala propia de los vuelos de Levy. Esta idea ha 
sido implementada en t\'ecnicas de optimizaci\'on computacional \cite{Computacion}.

\section{Datos sobre mutaciones puntuales en el EELP}\label{s1}

\begin{figure}
\begin{center}
\includegraphics[width=8cm]{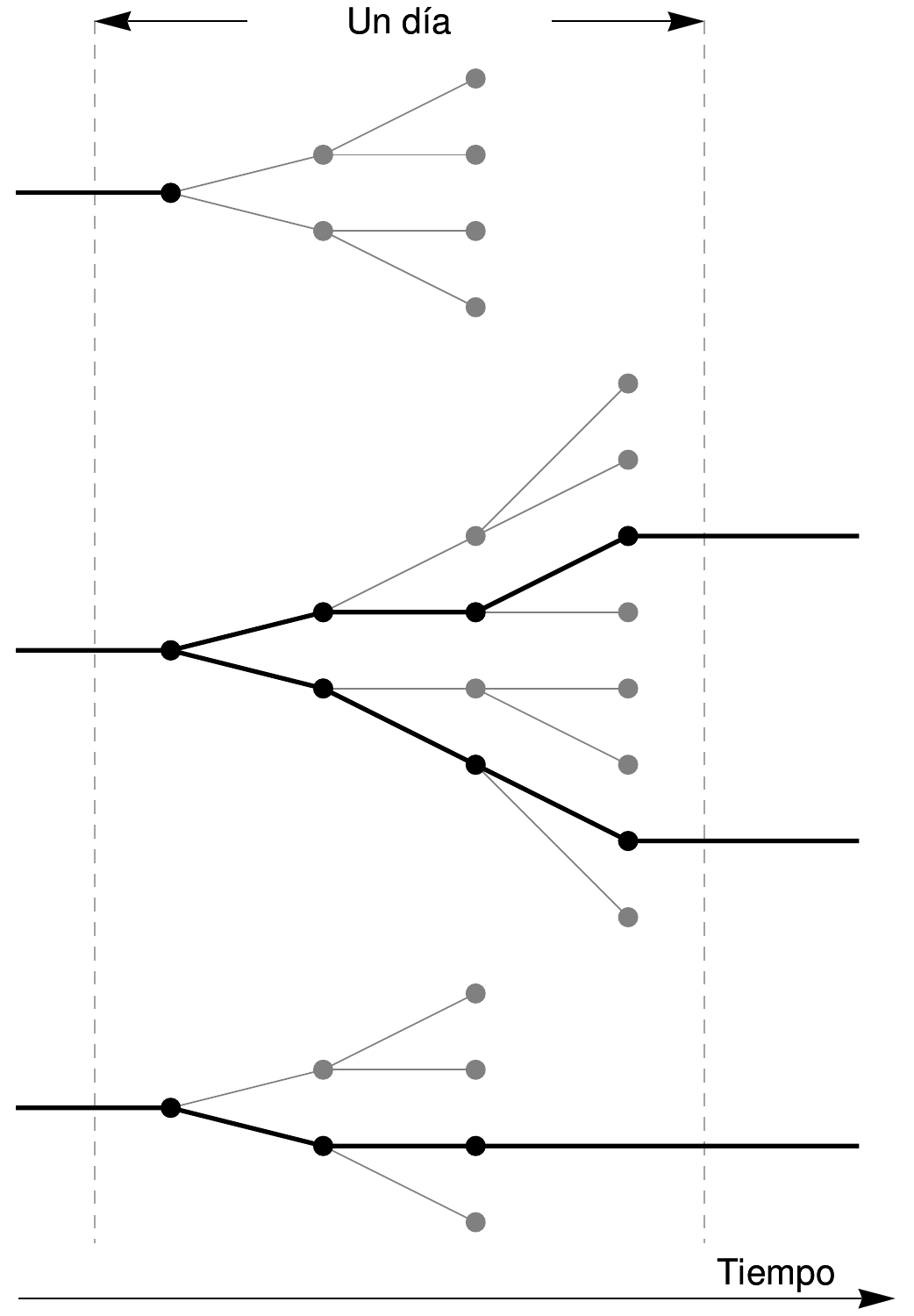}
\caption{ Representaci\'on esquemática de la filogenia de la evolución de las bacterias durante un d\'ia en el EELP.
Se han dibujado solo 2-3 divisiones celulares, cuando en el experimento se suceden 6-7.} \label{filoBact}
\end{center} 
\end{figure}

\begin{figure}
\begin{center}
\includegraphics[width=10cm]{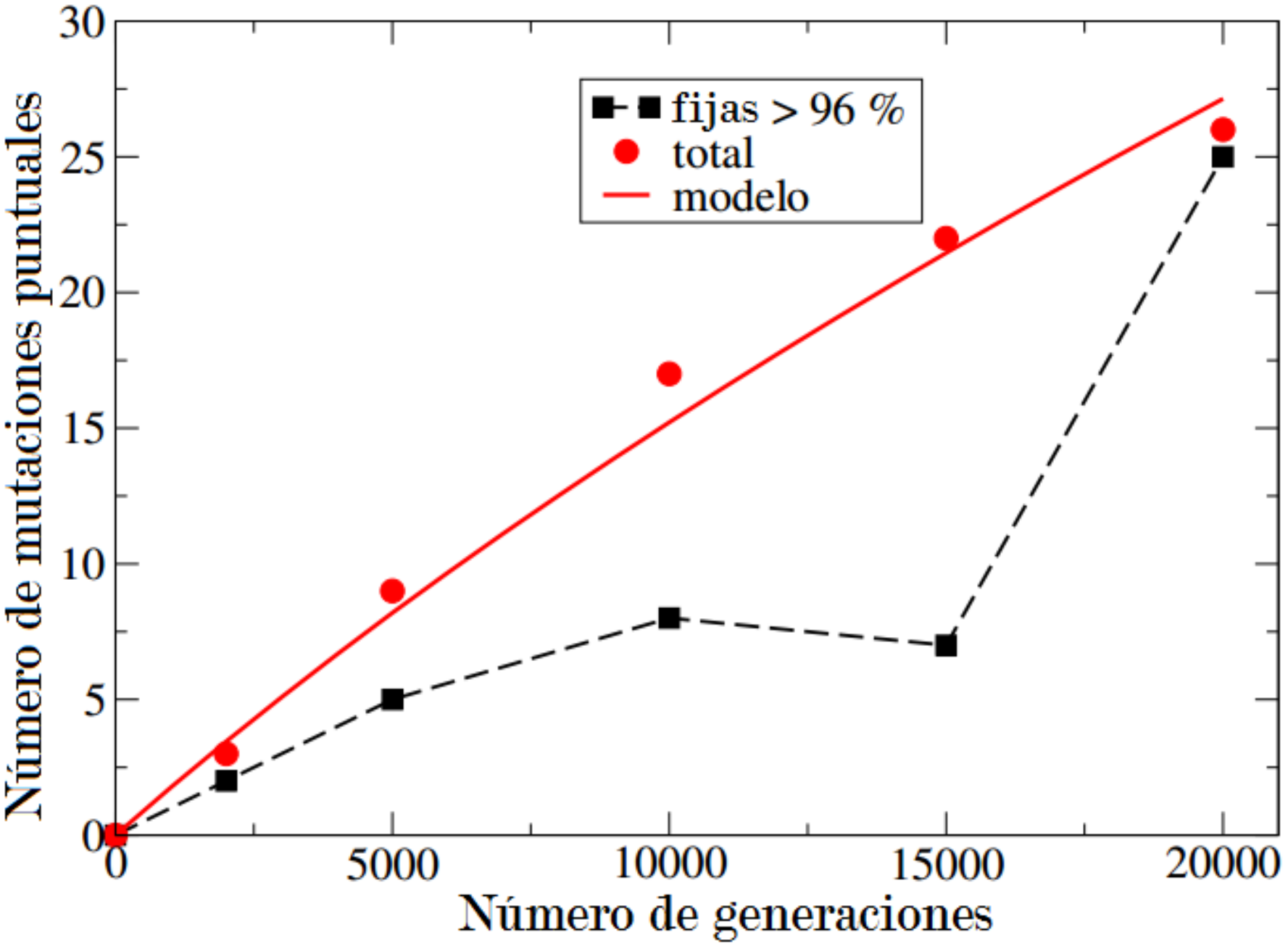}
\caption{ Cantidad de mutaciones puntuales como funci\'on del tiempo (n\'umero de generaciones) en la poblaci\'on Ara-1 del EELP \cite{A1}. 
Est\'an incluidos en la figura los datos desde la generación 0 (cadena ancestral, tomada como referencia) hasta la 20 000. } \label{A1}
\end{center} 
\end{figure}

Un d\'ia de evolución en el EELP se representa esquemáticamente en la \fig{filoBact} por trayectorias de filogenia. Usualmente los linajes con mutaciones neutras o 
deletéreas son truncados, mientras que las mutaciones beneficiosas les confieren ventajas evolutivas a los clones y por tanto, tienen mayor probabilidad 
de continuar. Una vez que aparecen, las mutaciones beneficiosas se fijan en más del $50\%$ de la población después de un tiempo de dominancia.  Hablando sin 
mucho rigor, si $P_b$ es la tasa de mutaciones beneficiosas en la población y $\tau_f$ el tiempo necesario para que se fije una, entonces la cantidad de 
mutaciones fijadas en un tiempo $t$ será aproximadamente $t/(\tau_b + \tau_f )$, donde  $\tau_b = 1/P_b$. En la \fig{A1} aparecen datos de mutaciones 
puntuales que fueron tomados de \cite{A1}. En ella se muestran los valores medidos en las generaciones 2 000, 5 000, 10 000, 15 000, 20 000, 30 000, 
y 40 000. Los otros dos puntos restantes no se incluyeron porque la aparición del fenotipo mutante en la generación 27 000 hace que la frecuencia de mutaciones
aumente en 100 veces. 

El experimento permite determinar frecuencias de mutaciones puntuales superiores al $4\%$ en la población. Los autores reportan mutaciones ``fijas'', 
lo que significa que sus frecuencias $f$, son mayores que el $96\%$, como también las mutaciones con: $4\%<f<96\%$. Los datos llamados ``fijas'' 
en la figura muestran un incremento lineal para tiempos pequeños con pendiente $1.0\times10^{-3}$ mutaciones/generación, 
lo cual brinda una estimación de $P_b$. Por otro lado, los datos 
llamados ``total'' son nuestras estimaciones para el n\'umero de mutaciones que se pueden detectar en un clon en total, no importa cuál sea. La pendiente en 
el incremento lineal inicial es un poco mayor: cerca de $1.8\times 10^{-3}$ mutaciones/generación. Debemos indicar, una vez más, que estos números son 
para la población completa. La tasa de mutaciones beneficiosas en un solo linaje es $p_b=P_b/N_{cel} \approx 0.8\times10^{-3} /(5\times10^6 ) \approx 
3\times10^{-10}$ mutaciones/generación. Notemos que en el experimento el número de linajes de células y el tamaño del genoma son 
aproximadamente el mismo ($5\times 10^6$), lo que a veces puede traer confusión. Notemos también que el número total de mutaciones muestra una dependencia sublinear 
para largos tiempos. Esto es una consecuencia del hecho de que $\tau_f$ usualmente se incrementa cuando se añade una nueva mutación beneficiosa sobre 
las existentes, fenómeno conocido como epistasis.  Para ajustar los datos en la figura usamos la dependencia 
del modelo propuesto en \cite{Adaptation_Lenski}: $2s(\sqrt{1 + aN_{gen}} -1)/a$, donde $s$ es la pendiente en $N_{gen} = 0$ y $a$ es un parámetro.

\section{Datos sobre mutaciones no locales en el EELP}\label{s2}

En la Ref. \cite{mBio} se proveen datos de reajustes cromosómicos grandes. De acuerdo a las limitaciones del experimento, los autores no pueden detectar 
reajustes menores que 5 kilo pares de bases (Kpb). Por otro lado, ellos solo pueden hacer mediciones en clones, que son representativos 
de una población, los cuales pueden exhibir grandes desviaciones de los valores medios. 

El primer grupo de resultados involucran una secuencia temporal de clones de la población Ara-1, como en la sección anterior. Esto es, 
muestras en las generaciones 2 000, 5 000, 10 000, 15 000, 20 000, 30 000, 40 000 y 50 000. Recordemos la parte espacial $\pi(l)$ de nuestro ansatz 
para la probabilidad de saltos largos. Nosotros no distinguimos entre los diferentes tipos de reajustes: delecciones, inserciones, translocaciones e inversiones.

\begin{figure}
\begin{center}
\includegraphics[width=10cm]{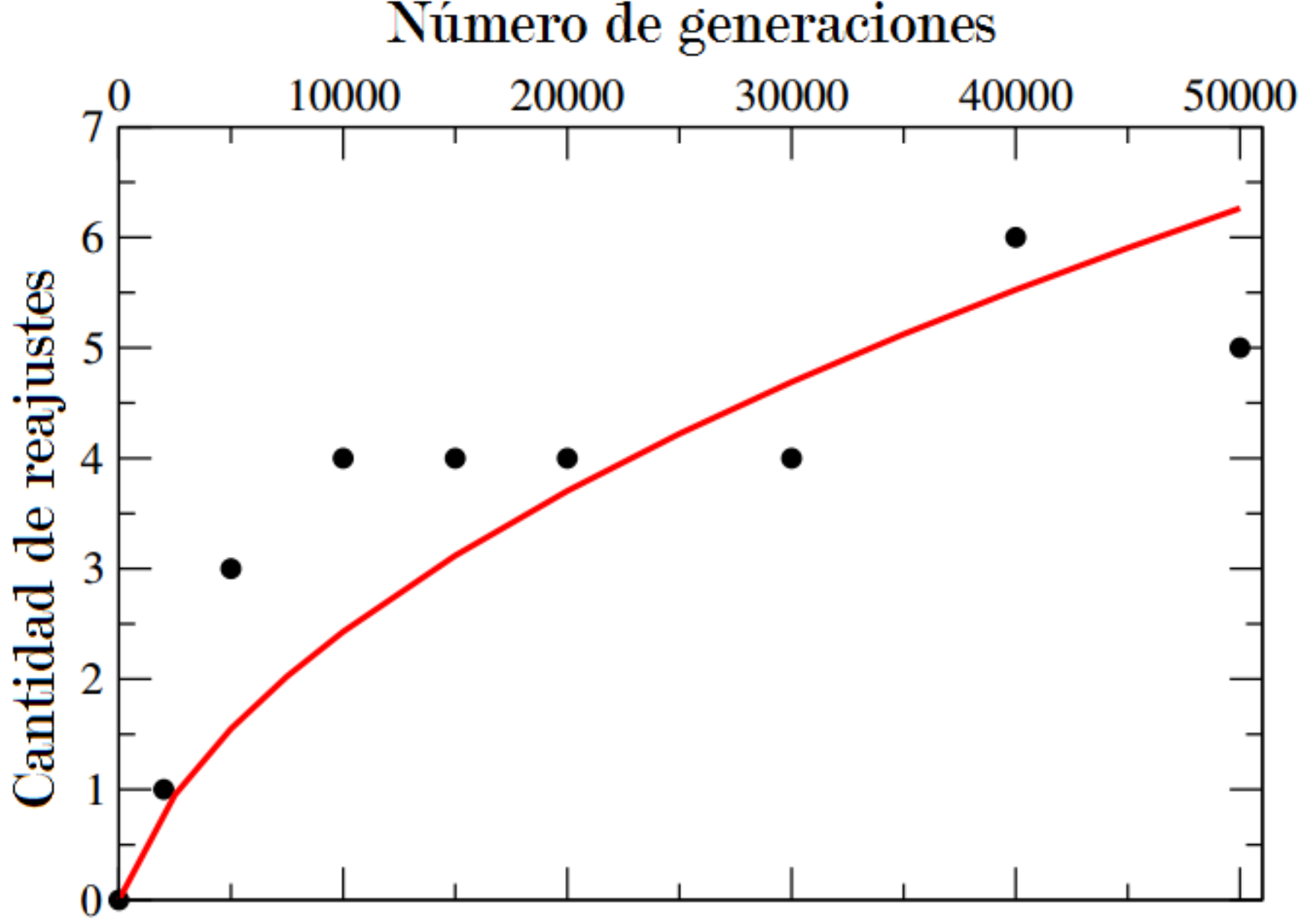}
\includegraphics[width=9.7cm]{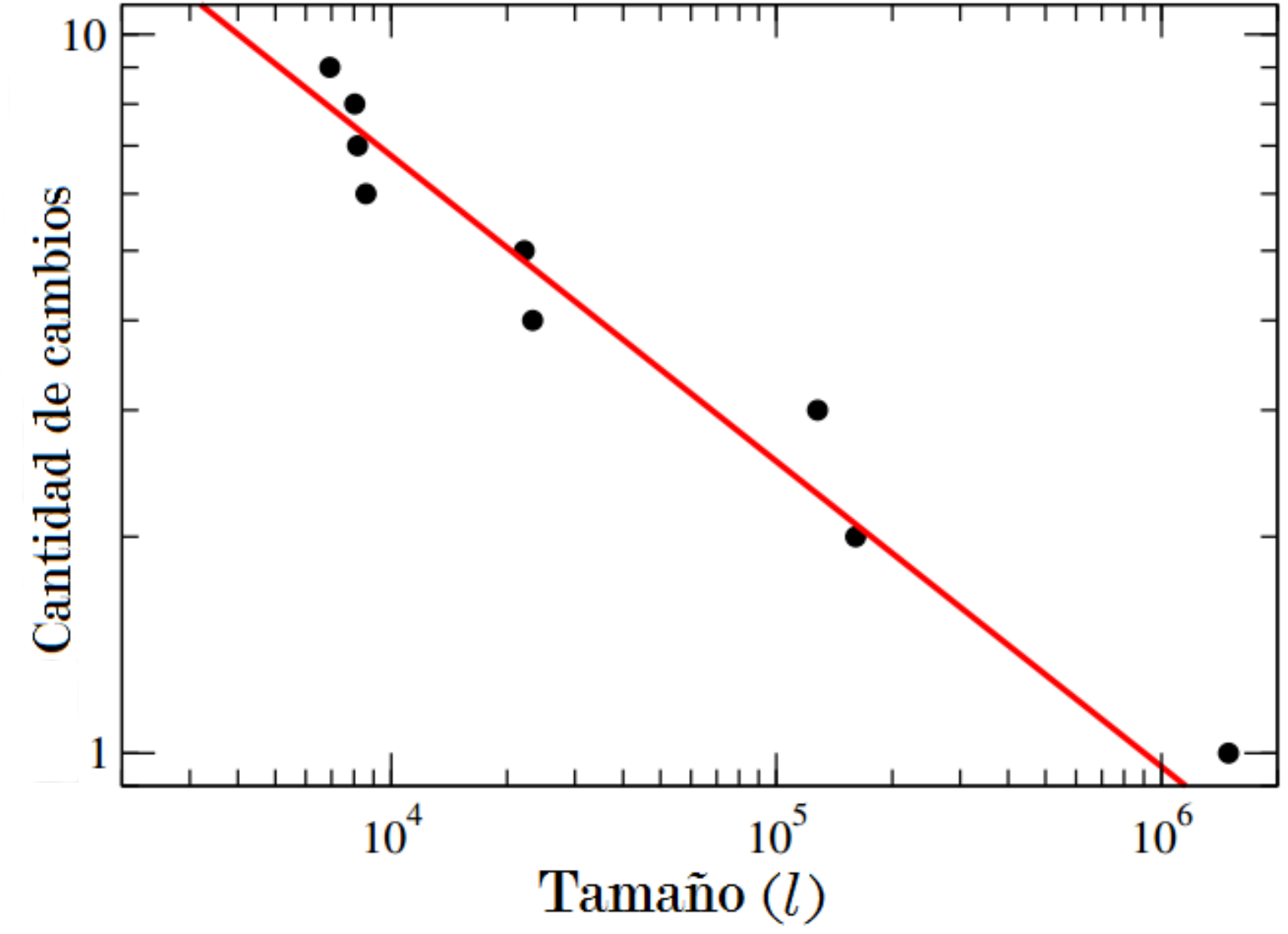}
\caption{  Arriba: N\'umero de reajustes cromosómicos grandes (mayores que 5 Kpb) en clones de la población Ara-1 como funci\'on del tiempo 
(n\'umero de generaciones). Abajo: Log-log plot de la distribuci\'on del tamaño de los eventos (ver explicaci\'on detallada en el texto).} \label{A2A3}
\end{center} 
\end{figure}

La \fig{A2A3} de arriba muestra los números de eventos detectados como función del tiempo (número de generaciones). La mayoría de los reajustes parece 
que se fijan, en el sentido de son detectados también en tiempos posteriores. Para ajustar los datos usamos la misma función que 
para las mutaciones puntuales: $2s(\sqrt{1 + aN_{gen}} -1)/a$. A partir de la pendiente obtuvimos una estimación aproximada para la tasa de cambios 
beneficiosos en la población: $P_{bSL} \approx 5\times 10^{-4}$ cambios/generación. Para un solo linaje de c\'elulas, $p_{bSL} \approx 10^{-10}$
cambios/generación.

La \fig{A2A3} de abajo, por otro lado, refleja la estadística del tamaño. Usamos una gráfica log-log con el tamaño $l$ en el eje de las abscisas y el número 
de reajustes con tamaños mayores o iguales a $l$ en el de las ordenadas. De acuerdo con nuestro ansatz, este número es igual a: 

\begin{equation}
C(\nu -1)\int_l^{\infty} \frac{dx} {x^\nu} = \frac{C} {l^{\nu -1}}
\end{equation}

\noindent
donde C es una constante de normalización. Notemos que cuando un cambio aparece en un tiempo dado y se fija, nosotros no lo contamos como un 
evento diferente en un tiempo posterior.  

Los datos de la \fig{A2A3} se ajustan bastante bien por la función $C/l^{\nu-1}$, con $\nu = 1.42$. M\'as abajo consideraremos más datos con una mejor estadística. 

\begin{figure}
\begin{center}
\includegraphics[width=10cm]{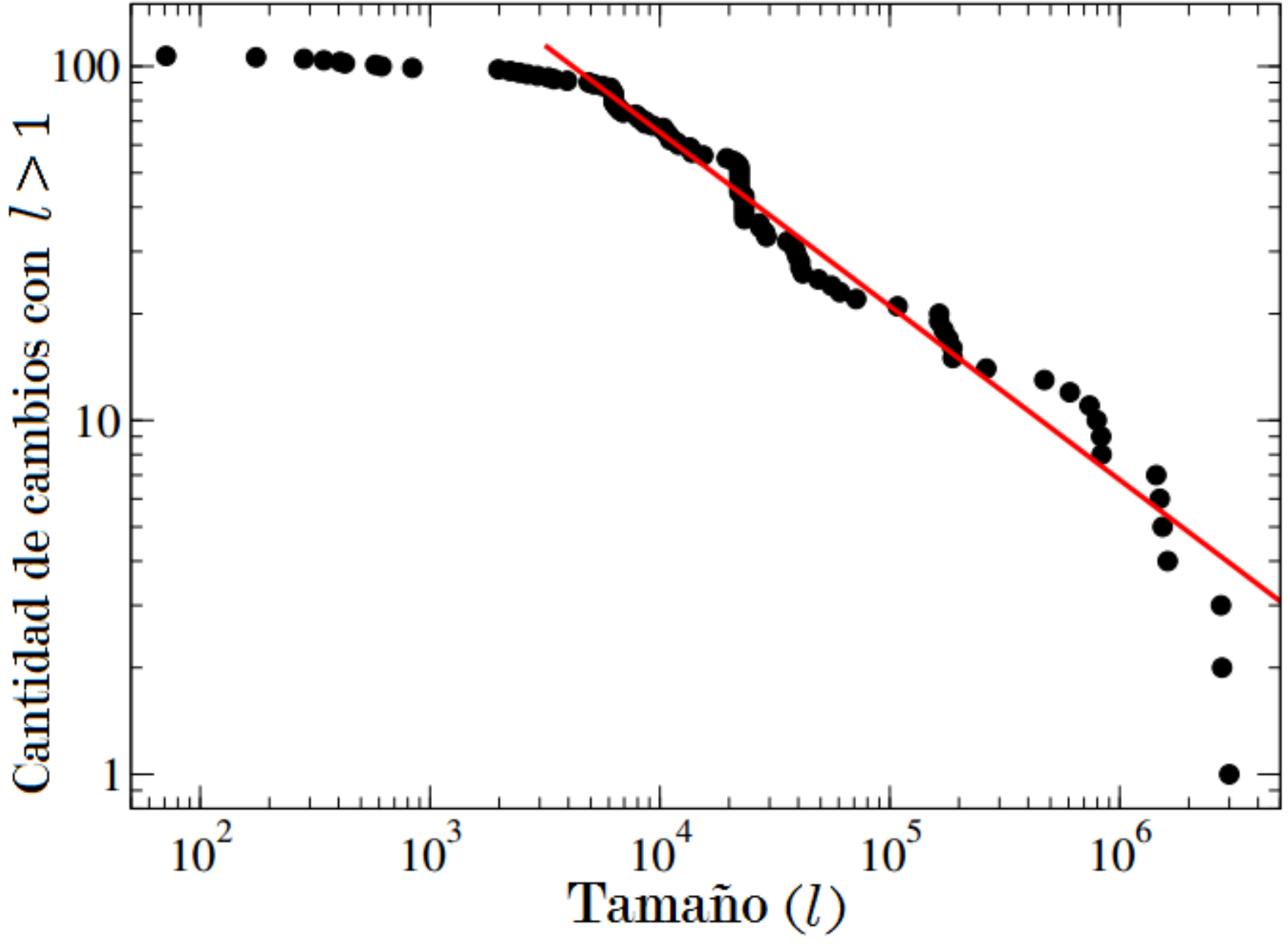}
\caption{ Log-log plot de la distribuci\'on del tamaño de grandes reajustes en clones obtenidos de las 12 poblaciones que evolucionan independientemente
en el EELP, muestreados en la generaci\'on 40 000. La l\'inea es un ajuste con $\nu=1.49$.} \label{A4}
\end{center} 
\end{figure}

El segundo grupo de datos proviene de clones cosechados de las 12 poblaciones que evolucionan independientemente en el experimento, muestreados en 
la generación 40 000. Existen 110 reajustes grandes en dichos clones. Los resultados, mostrados en la \fig{A4}, son muy bien fiteados por la dependencia 
$C/l^{\nu-1}$ , con $\nu = 1.49$, sugiriendo un valor límite del exponente  $\nu = 3/2$. La pendiente cambia para $l < 5$ Kpb porque el experimento no puede 
detectar todos los cambios que ocurren con esos valores de $l$, como se mencionó anteriormente.

En resumen, en las secciones \ref{s1} y \ref{s2} hemos mostrado que las mutaciones en el EELP pueden describirse como trayectorias de Levy. También hemos determinado
las frecuencias de mutaciones puntuales y no locales, as\'i como el exponente de Levy, $\nu$.

\section{Descripción del {\it \textbf{fitness}} en el EELP}

Para definir el parámetro de {\it fitness}, seguiremos un esquema parecido al de la Ref \cite{Notes_Fitness}. En el caso del EELP,  una población inicial consiste en 
individuos con genotipos $z_0$ (cepa ancestral de {\it E. Coli}) y las réplicas siempre tienen el mismo genotipo que sus padres (reproducción por mitosis), salvo en 
caso de mutaciones, donde se producen genotipos $z_i$ con algunas variaciones. Asumamos que luego de cierta cantidad de replicaciones existe una 
distribución de genotipos en la población. Si del tipo $z_i$, existen $N(z_i)$ individuos, los cuales sobreviven con probabilidad 
$l(z_i)$ y dan lugar a $m(z_i)$ réplicas, entonces la cantidad de individuos nuevos que se obtienen a partir de cada tipo en la siguiente generación será:

\begin{equation}
N'(z_i)=l(z_i)m(z_i)N(z_i)\equiv W(z_i)N(z_i)
\end{equation}

\noindent
Notemos que el factor $W$ determina las frecuencias relativas de los genotipos y por tanto, constituye una completa descripción del {\it fitness}. 
Visto de esa forma, la competencia entre los genotipos estará determinada por la razón entre sus {\it fitness}, por tanto, podemos normalizar el 
{\it fitness} al de uno de los genotipos, digamos $z_0$:

\begin{equation}
N'(z_i)= w(z_i)N(z_i), \label{n_ind}
\end{equation}

\noindent
donde $w(z_i)=W(z_i)/W(z_0)$ y $w(z_0)=1$

El número total de individuos de la siguiente generación será exactamente:

\begin{equation}
N'= N\sum _{i} w(z_i)f(z_i), 
\end{equation}

\noindent
donde $N$ es la cantidad de individuos en la generación anterior y $f(z_i)$ la frecuencia relativa de cada genotipo en la generación anterior 
($f(z_i)=N(z_i)/N$). Como las mutaciones son eventos poco probables, s\'olo una pequeña fracci\'on del total de los nuevos individuos tendr\'an 
genotipos diferentes. Notemos que la aparición de nuevos genotipos en un linaje estar\'a determinada por la supervivencia del individuo de la 
generaci\'on anterior a partir del cual se genera, o sea, se rige por el {\it fitness} de su antepasado.

Si realizamos replicaciones sucesivas de la población a partir de la ecuaci\'on (\ref{n_ind}), obtendremos una aproximación de la cantidad de 
individuos en la generaci\'on $t$:

\begin{equation}
N^t(z_i)= w(z_i)^tN_0(z_i) \label{replic}
\end{equation}

En una situación hipotética donde existe un genotipo mayoritario en la población, como al inicio del EELP, la frecuencia del mismo ser\'a pr\'acticamente
la unidad y la cantidad de individuos con ese genotipo en la siguiente generaci\'on no tendr\'a cambios apreciables ($N'(z_0) \approx N(z_0)$). 
Analicemos ahora un genotipo singular $z_x$ de la poblaci\'on, pobremente representado por un pequeño n\'umero de individuos.
En el caso extremo de s\'olo existir un individuo con ese genotipo, el n\'umero de individuos con dicho genotipo en las generaciones siguientes 
estar\'a determinado \'unicamente por su {\it fitness} (ecuaci\'on (\ref{replic}) con $N_0=1$). As\'i, un individuo con una mutación 
beneficiosa que elevó su {\it fitness} de $1$ a $1.1$ se reproducir\'a $1.1$ veces m\'as r\'apido y en alg\'un momento se har\'a mayoritario. 

En el EELP se controla la cantidad de individuos de una población, manteni\'endose un valor de $5\times10^6$ linajes celulares constante en el tiempo. 
Si tenemos esto en
cuenta en nuestro ejemplo anterior, un estimado sencillo de la cantidad de generaciones o el tiempo de dominancia, $t_d$, necesario para que un cierto 
genotipo con {\it fitness} $w_1$ se haga mayoritario frente a otro  de {\it fitness} $w_2<w_1$ es:

\begin{equation} \label{domina}
{(w_1/w_2)}^{t_d}=N_f,
\end{equation}

\noindent
donde $N_f$ es una fracci\'on de la poblaci\'on, por ejemplo la mitad. 
Cuando existen varios genotipos compitiendo entre s\'i, la dominancia del de mayor {\it fitness} no es tan trivial: se alcanza cuando la cantidad de 
individuos con ese genotipo supere a la del resto. La fracci\'on de la poblaci\'on que representa dicha condici\'on depende de los valores de {\it fitness}
de cada genotipo en la competencia, este fenómeno es conocido como interferencia clonal. 

\section{Simulando las mutaciones + la selección natural en el EELP}

Con el objetivo de comprobar nuestras hip\'otesis sobre las mutaciones, simularemos la dinámica del EELP, representada en la \fig{filoBact}. 
Como explicamos en el Cap\'itulo \ref{cap1}, en el experimento se parte de un cultivo de $5\times10^6$ bacterias, cada bacteria se reproduce
durante 6 o 7 generaciones en promedio alcanz\'andose 100 veces la cantidad inicial. De las 500 millones de bacterias resultantes en un d\'ia
de expansi\'on clonal, se escoge entonces el $1\%$, obteni\'endose la cantidad inicial. Como a lo largo del experimento el n\'umero de linajes celulares 
permanece constante e igual al tamaño inicial del cultivo, por una cuesti\'on de costo computacional, en la simulación nos restringimos a describir una 
duplicaci\'on de las 5 millones de trayectorias seguidas en el EELP. 

Estamos interesados en la descripción temporal de las mutaciones, por lo que encontramos conveniente seguir la evolución del {\it fitness} de cada 
una de las trayectorias. Los primeros pa\-rá\-me\-tros a tener en cuenta son las probabilidades por c\'elula de que aparezca una mutación y de que esta sea 
beneficiosa. Sus valores num\'ericos para la población Ara-1 fueron estimados pre\-via\-men\-te en las secciones anteriores a partir de los datos del experimento. 
$p_{mut}$ indica cu\'ando hay mutación y $p_b$ indica cu\'ales de ellas son beneficiosas. Cuando una c\'elula experimenta una mutación beneficiosa se produce 
un incremento de su {\it fitness}. Otros casos de mutaciones son las neutrales, las cuales a pesar de modificar al ADN, no producen ning\'un efecto 
sobre el {\it fitness}. La \'ultima posibilidad de mutaciones es que sean perjudiciales, que simulamos con la misma probabilidad de las neutrales 
y disminuye el {\it fitness} 
hasta en un $70 \%$, este par\'ametro es arbitrario en principio, pero constituye un valor de corte a partir del cual consideramos despreciable la probabilidad 
de que una c\'elula con un {\it fitness} tan reducido sobreviva durante 7 generaciones (un d\'ia) y logre ser detectada en el EELP.

Para incrementar el {\it fitness} en caso de mutación beneficiosa utilizamos el modelo descrito en \cite{Adaptation_Lenski}, donde la ventaja $s$ de
la mutaci\'on se distribuye exponencialmente con densidad de probabilidad $\alpha e^{-\alpha s}$. La ventaja se define a partir de los valores 
de {\it fitness} antes y después de la mutación como: $w'=w(1+s)$. El par\'ametro $\alpha$ de la distribución también cambia en el tiempo: 
$\alpha'=\alpha(1+gs)$ (s\'olo en caso de mutación beneficiosa), donde $g$ se ajusta para cada población y vale $g=4.035$ para la población
Ara-1 \cite{Adaptation_Lenski}.


Mencionamos que para simular la selección natural no podemos hacer una expansi\'on clonal de 5 a 500 millones de c\'elulas, como en el experimento, y después escoger 
aleatoreamente 5 millones. Como debemos mantener arreglos de {\it fitness}, n\'umero de mutaciones, etc. las dimensiones se hacen muy grandes.

En vez de eso, hacemos una sola duplicaci\'on y escogemos la mitad de las c\'elulas. El algoritmo en cuesti\'on es as\'i:

\begin{itemize}
 \item El mayor {\it fitness} se toma como referencia, $w_{max}$. 
  \item De cada clon se hacen dos copias y se determina si existen mutaciones con probabilidad condicional $p_{mut}=10^{-3}$ y, de ellas, si son
  beneficiosas, con probabilidad $p_b=10^{-6}$. Estos valores son dictados por el experimento.  
  \item La permanencia, o no de las copias que provienen de $w_{max}$ se hace aleatoreamente con probabilidad $1/2$. Para el resto de los clones la 
  probabilidad es $w/(2w_{max})$.
\end{itemize}

A veces, después de este procedimiento no se han completado los 5 millones de c\'elulas, faltan unos miles. Ellas se escogen aleatoreamente ya que,
como son tan pocas, las frecuencias relativas no tienen un gran peso.

\begin{figure}
\begin{center}
\includegraphics[width=10cm]{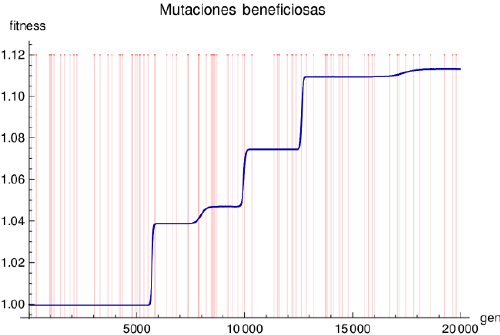}
\includegraphics[width=10cm]{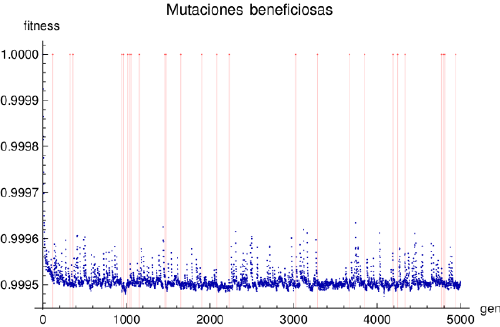}
\includegraphics[width=10cm]{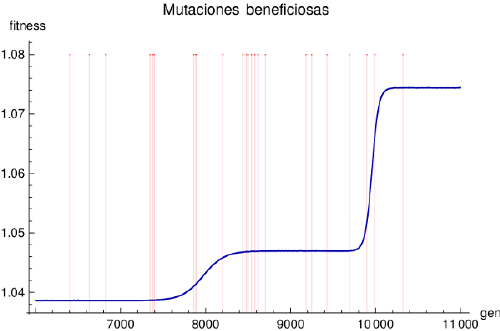}
\caption{ Historia del {\it fitness}  medio en la simulación hasta 20 000 generaciones y detalles en distintas etapas.
Las l\'ineas verticales representan la aparición de mutaciones beneficiosas, la mayor\'ia de las cuales no logra imponerse.} \label{FHD}
\end{center} 
\end{figure}

\begin{figure}
\begin{center}
\includegraphics[width=10cm]{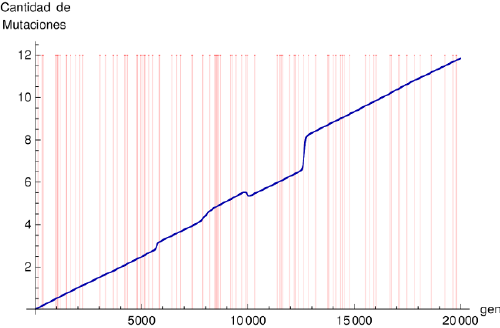}
\includegraphics[width=10cm]{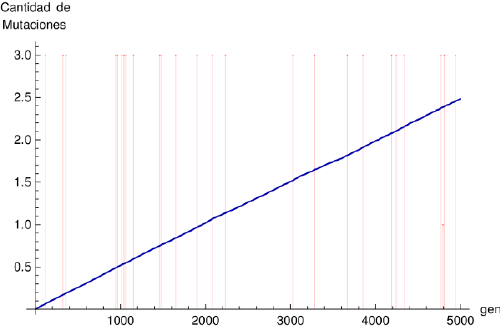}
\includegraphics[width=10cm]{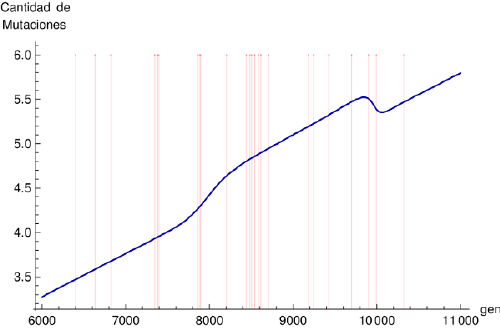}
\caption{ Historia de la cantidad media de mutaciones en la simulación hasta 20 000 generaciones.} \label{SHD}
\end{center} 
\end{figure}

En la \fig{FHD} se muestra la evolución temporal del {\it fitness} promedio de la población a lo largo de 20 000 generaciones, obtenida de la simulación.
El comportamiento es similar al de la Ref.\cite{Adaptation_Lenski}. También se reproduce el comportamiento de la evolución temporal de la cantidad 
de mutaciones, análogo a la \fig{A1} y mostrado en la \fig{SHD}. En las gráficas, las generaciones donde aparecen mutaciones beneficiosas están marcadas. 
Los escalones de ambas gráficas indican que cierta mutación beneficiosa se fijó en la población y elevó el {\it fitness} medio abruptamente. Otras de ellas 
fracasan en la competencia por sobrevivir, sin embargo, constituyen fuentes de variabilidad genot\'ipica.

También podemos analizar el tiempo que necesita una mutación beneficiosa para fijarse en la población. Resulta que mientras 
m\'as beneficiosas sean, es decir, mayor {\it fitness} tenga la célula mutada, m\'as rápido \'esta se propagar\'a. 

\begin{figure}
\begin{center}
\includegraphics[width=15cm]{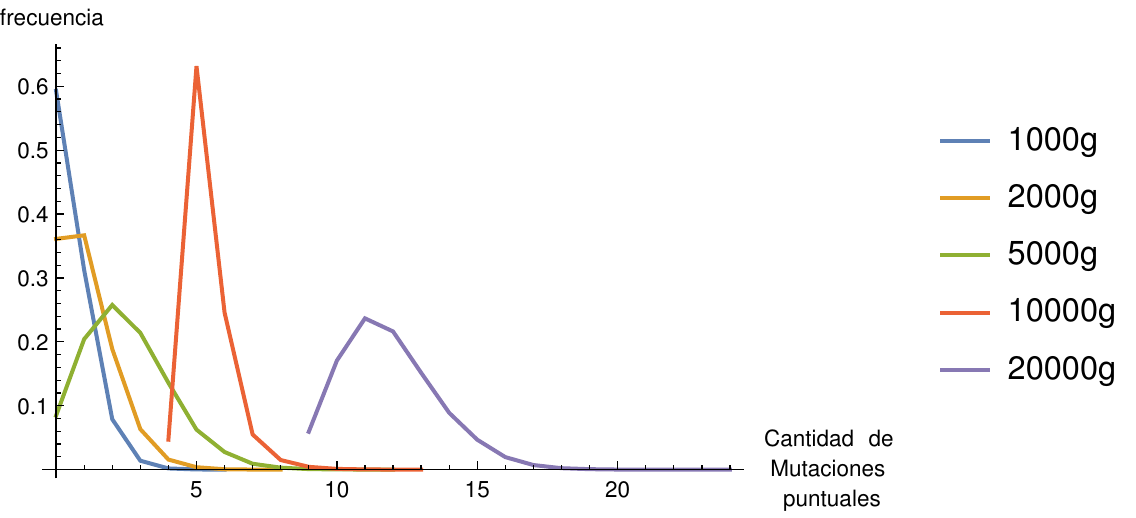}
\caption{ Perfiles de la cantidad media de mutaciones en distintos momentos la simulación: en las generaciones 1 000, 2 000, 5 000, 10 000 y 20 000.} \label{FMD}
\end{center} 
\end{figure}

En la \fig{FMD} se muestran datos recogidos en distintas generaciones de la cantidad de mutaciones producidas. Las curvas que se obtienen son consistentes 
con una distribución poissónica truncada: es decir, de la forma $Ne^{-\lambda}\lambda^k/k!$; donde $k$ es variable discreta inicializada en cierto valor positivo,
$\lambda$ es un par\'ametro que resulta el valor medio de la distribuci\'on y $N$ es una constante de normalizaci\'on. 
A partir de la 
generación 10 000 ya todas las trayectorias tiene al menos 4 mutaciones, mientras que en la generación 20 000 tienen al menos 9. Contando la cantidad de 
escalones en la gr\'afica del {\it fitness} medio sabemos que al menos 5 de esas mutaciones son beneficiosas. Otro resultado interesante es referente a 
la variabilidad genot\'ipica. Resulta que, en los primeros instantes de la simulaci\'on, la mayor\'ia de las c\'elulas tienen {\it fitness} unitario y existe
poca variabilidad debido a que no le han dado tiempo de propagarse las mutaciones. A medida que avanza la simulaci\'on el abanico de posibilidades aumenta,
lo que se traduce en un aumento del ancho de la distribución de las mutaciones (ge\-ne\-ra\-cio\-nes 1 000, 2 000 y 5 000). A veces el abanico se cierra instantes
posteriores a la aparición de una mutaci\'on beneficiosa en un genotipo, que se vuelve dominante y extingue a la mayor\'ia de los dem\'as competidores 
(ge\-ne\-ra\-ci\'on 10 000). Esto constituye una manifestaci\'on de la interferencia clonal. En la generaci\'on 20 000 se vuelve a ensanchar la distribución despu\'es
de un per\'iodo de evoluci\'on sin que se fije alguna mutaci\'on beneficiosa. 

\begin{figure}
\begin{center}
\includegraphics[width=10cm]{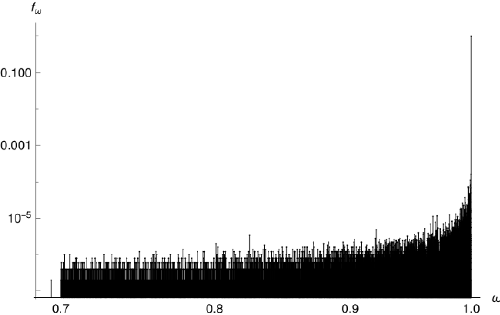}
\includegraphics[width=10cm]{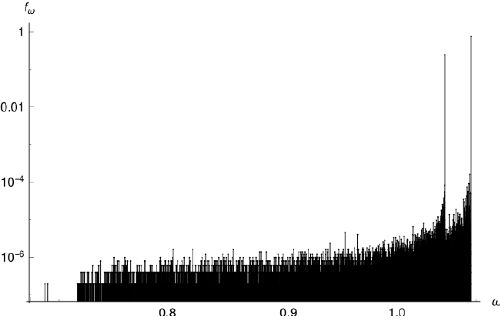}
\includegraphics[width=10cm]{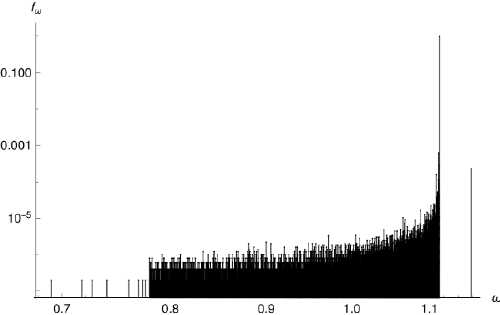}
\caption{ Tres tipos de situaciones que se pueden dar con el {\it fitness} en el EELP y la simulación, referentes a la interferencia clonal.
En la figura de arriba predomina un s\'olo genotipo. En la figura del medio compiten dos genotipos, pero el de mayor {\it fitness} es el dominante. 
En la figura de abajo tambi\'en compiten dos genotipos, pero el de mayor {\it fitness} no se ha interpuesto.} \label{fitD}
\end{center} 
\end{figure}

Por otro lado, el perfil de frecuencias de
todos los genotipos obtenidos en varios instantes de la simulaci\'on son mostrados en la \fig{fitD}. Se escogieron tres situaciones diferentes que
se pueden dar y que ocurrieron en el perfil de las generaciones 2 000, 10 000 y 20 000. En la primera tenemos que la mayoría de las c\'elulas tienen
un {\it fitness} de 1, que se corresponde con el m\'aximo valor fijado en la poblaci\'on. En la segunda se observan dos picos de frecuencia correspondientes
a mutaciones beneficiosas en clones que compiten entre s\'i, la de mayor {\it fitness} es mayoritaria y posteriormente puede imponerse (interferencia clonal).
En el tercer caso tenemos otra interferencia clonal donde el mayor {\it fitness} no se ha hecho mayoritario.

Algunas diferencias de nuestra simulación con respecto al experimento, son que subestimamos los máximos de {\it fitness} y la cantidad de mutaciones en 
promedio. La razón de este resultado podr\'ia ser, que la probabilidad de ocurrencia de mutaciones se estima suponiendo que, en el EELP, la 
mayoría de la mutaciones beneficiosas producidas se fijaban en toda la población, contrario a lo que se observa en nuestra simulación. Posiblemente, 
las pro\-ba\-bi\-li\-da\-des reales de mutaciones sean m\'as altas y la fracci\'on de mutaciones beneficiosas que se pierden sea también tan alta, como en la
simulación.



\section{Aparición del genotipo mutante en el EELP}

\begin{figure}
\begin{center}
\includegraphics[width=12cm]{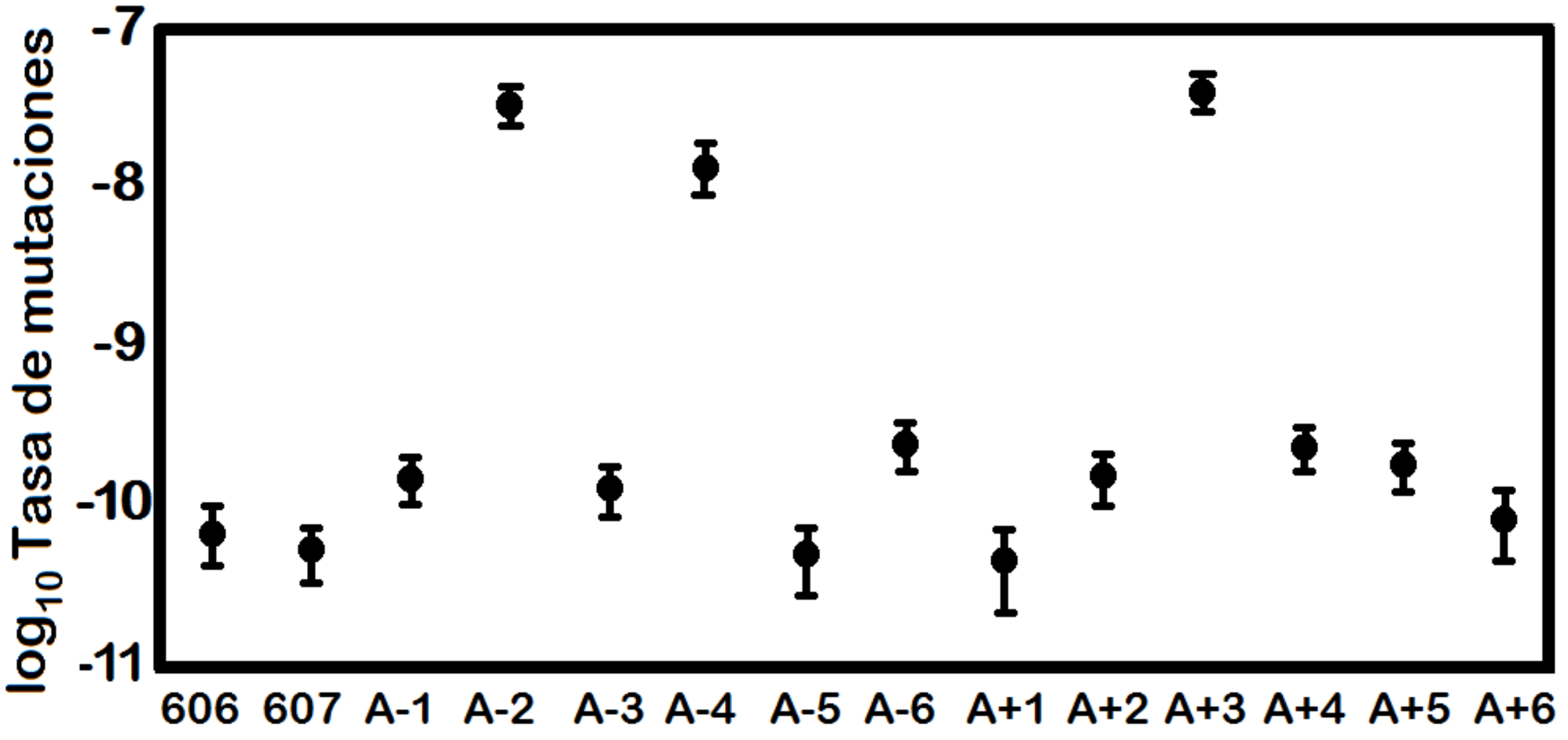}
\caption{Tres de las 12 poblaciones de {\it E. Coli} del EELP evolucionaron a fenotipos hipermutantes despu\'es de 10 000 generaciones.
Los datos muestran las tasas de mutaci\'on de las 12 poblaciones (Ara-1-Ara-6, Ara+1-Ara+6) y sus respectivos ancestros (606 y 607) al 
resistirse a un antibi\'otico. La poblaciones hipermutantes tienen cerca de 100 veces la tasa de mutaci\'on de sus ancestros, as\'i como
de las otras l\'ineas de evolución.} \label{FigLenski}
\end{center}
\end{figure}

En varias poblaciones del EELP aparece un ``mutante'', es decir, un genotipo donde la frecuencia de mutaciones aumenta aproximadamente 100 veces 
respecto a la cepa ancestral \cite{20mil_Lenski}. En la \fig{FigLenski},tomada de la Ref \cite{20mil_Lenski}, se muestra la frecuencia de mutaciones ante 
un antibiótico de las diferentes 
poblaciones del EELP, cuando se alcanza un n\'umero de generaciones $N_{gen}=10000$. 

Se observan 3 poblaciones donde el mutante se ha impuesto como genotipo dominante. Esto significa que su {\it fitness} 
es superior al promedio de la población, 
es decir, constituye un tipo especial de mutación beneficiosa que da lugar a su vez a un aumento de la tasa de mutaciones. 
Estudios en pacientes con fibrosis quística en los pulmones, infectados por {\it P. aeruginosa} \cite{antibiotico}, indican que también un genotipo mutante puede 
estar relacionado con la resistencia de capas bacteriales a los antibióticos.

\begin{figure}
\begin{center}
\includegraphics[width=10cm]{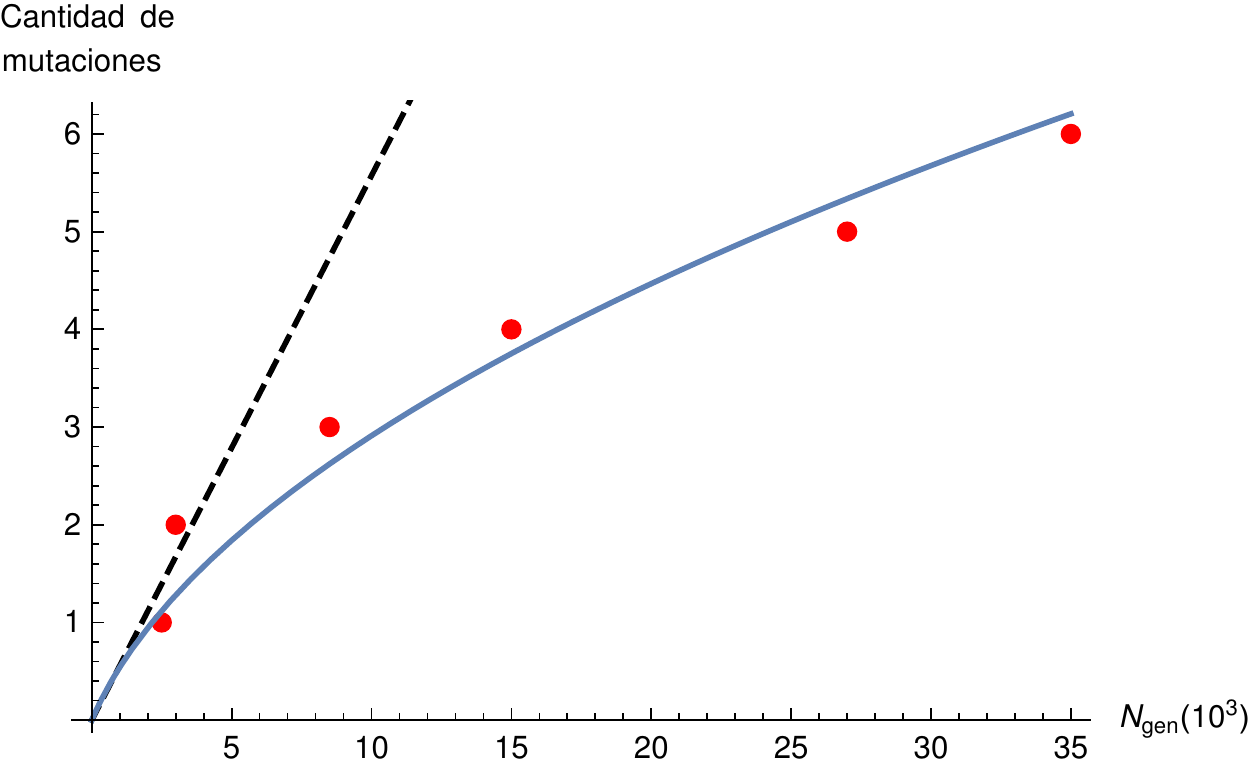}
\caption{Aparici\'on del genotipo mutante en las poblaciones del EELP. La gr\'afica muestra la cantidad de poblaciones en las que aparece como funci\'on
del n\'umero de generaciones. La l\'inea continua es un ajuste de los datos con el modelo dado en la Ref
\cite{Adaptation_Lenski}. La pendiente cerca del cero, representada con l\'ineas discontinuas, 
nos da un aproximado de su frecuencia de 
aparición: $5\times10^{-4}$ mutantes/generación.} \label{mutante}
\end{center}
\end{figure}

En la \fig{mutante} mostramos los datos del EELP sobre cuando aparece el genotipo mutante \cite{Adaptation_Lenski}. Los resultados son, de alguna forma, similares a la 
\fig{A1} con el número de mutaciones beneficiosas como función del tiempo. La pendiente cerca del cero nos permite apreciar una frecuencia aproximada de
$5\times10^{-4}$ para la aparición del mutante en una población. Para convertirse en mayoritario, el mutante debe sobrepasar la selección aleatoria y 
competir con otros clones, cuyos {\it fitness} aumentan con el tiempo, aumentando el tiempo de dominancia.

Dividiendo la frecuencia por el número de líneas de evolución celular ($5\times10^6$) en el experimento, obtenemos la probabilidad de aparición del 
mutante por célula: $p_{mut}=10^{-10}/cel$, aproximadamente la mitad de la probabilidad de aparición de una mutación beneficiosa.

Veamos que también pudiera existir alguna similitud con el cáncer.

\section{EELP y cáncer}\label{Xx}

Existe una gran diferencia entre la filogenesis de evolución en el EELP y en los tejidos humanos \cite{Frank}, que se muestra 
en la \fig{FilogeniaCancer}:

\begin{figure}
\begin{center}
\includegraphics[width=9cm]{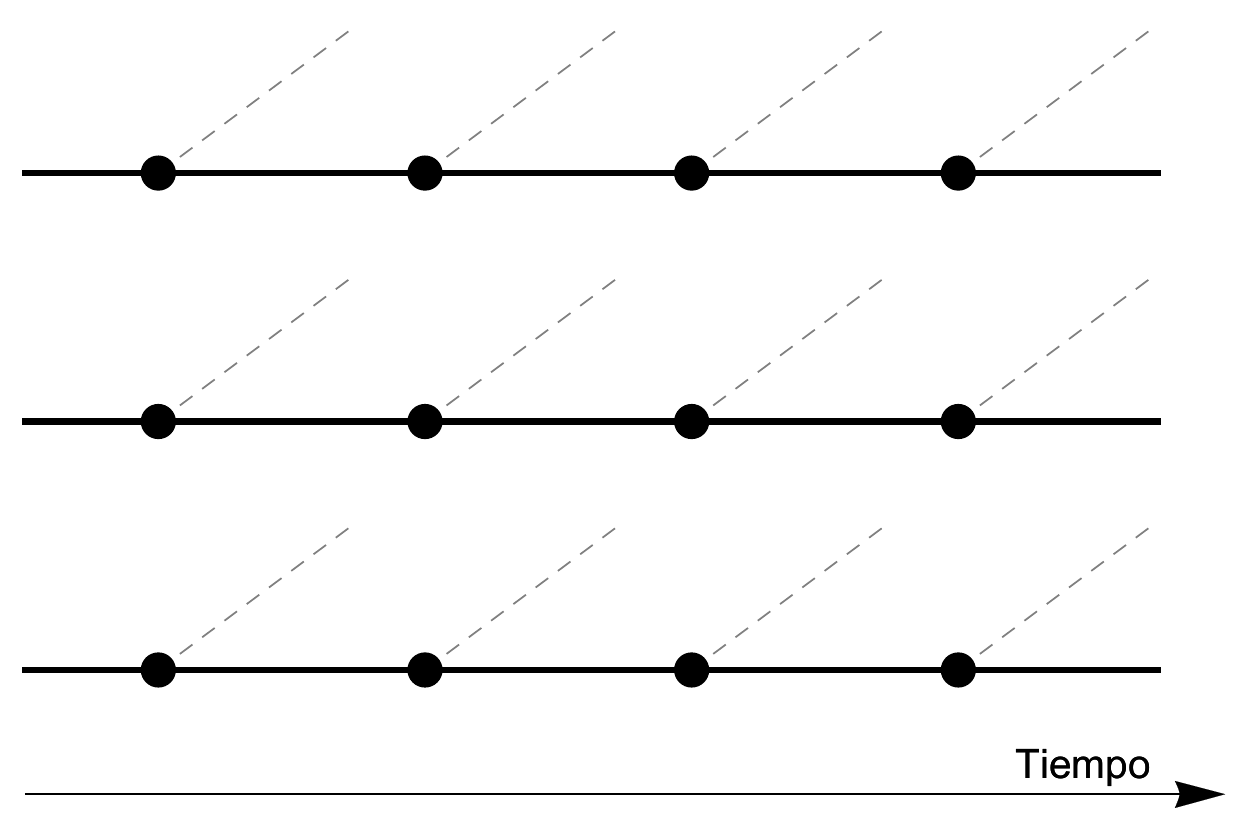}
\caption{Filogenia de las c\'elulas madres de los tejidos en la fase de homeostasis, las \'unicas capaces de acumular mutaciones. Las líneas de división celular, 
representadas discontinuamente, dan lugar a células especializadas de corta vida.} \label{FilogeniaCancer}
\end{center}
\end{figure}

La figura contiene s\'olo las líneas de evolución de las células madres en la fase de homeostasis. Las líneas de división celular, representadas
discontinuamente, dan lugar a células especializadas de corta 
vida. Las líneas de células madres son las \'unicas que perduran durante la vida del organismo y acumulan mutaciones, por eso se piensa que son ellas 
las pueden dar lugar al cáncer.

La diferencia fundamental de esta figura con la \fig{filoBact} es que no existe interferencia clonal (excepto en la fase de crecimiento del tejido), es decir, 
las líneas evolucionan independientemente. Esto pudiera ser una estrategia evolutiva para reducir la propagación de cáncer en el tejido, al igual que 
la división asimétrica del ADN \cite{Frank}, o sea que las células madres retienen el hilo ancestral del ADN al replicarse el mismo. La 
independencia de las líneas se refuerza a veces con la compartimentación en el tejido: una célula madre puede estar relacionada con una zona espec\'ifica en
el tejido, como sucede con las criptas intestinales.

Como similitud con el EELP podemos resaltar que el cáncer, en un gran porciento de los casos, es un genotipo mutante \cite{Frank}, lo que le confiere la 
habilidad para evadir la respuesta inmune.

\begin{figure}
\begin{center}
\includegraphics[width=11cm]{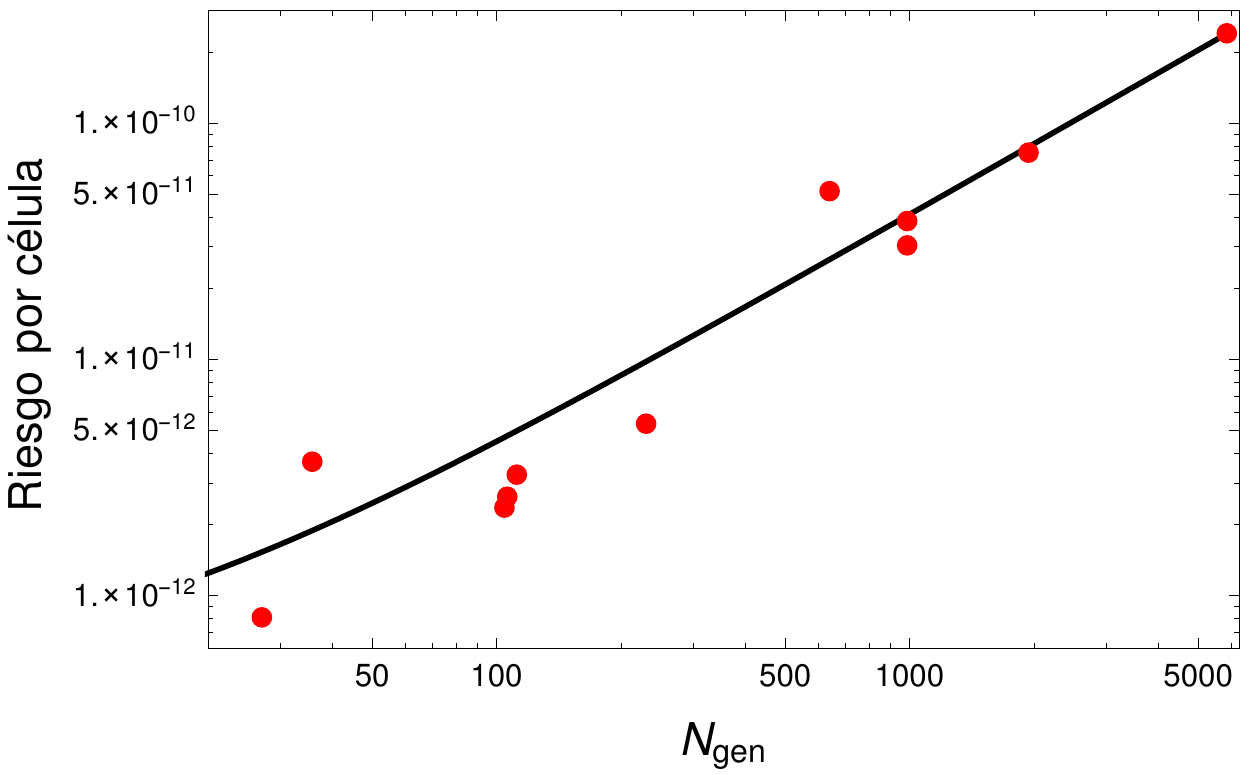}
\caption{Log-Log plot de los datos de riesgo de c\'ancer \cite{Tomasetti} por c\'elula en tejidos  que pensamos que est\'an poco afectados por otros 
factores carcinog\'enicos,
en funci\'on del n\'umero de generaciones de c\'elulas madres de cada tejido.
} \label{PTomasetti}
\end{center}
\end{figure}

En la \fig{PTomasetti} reploteamos los datos de Tomasetti y Vogelstein \cite{Tomasetti}, para un conjunto de tejidos que hemos denominado ``normales''. La diferencia es 
que el eje de las ordenadas ahora tiene el riesgo por célula y el eje de las abcisas, mide el número de generaciones a partir de la célula madre originaria 
que da lugar al tejido. La dependencia lineal es asombrosa a pesar de que los puntos representan órganos tan diversos como el cerebelo, la sangre y el colon, 
entre otros. Muestra lo que querían los autores: que en algunos tejidos el riesgo de cáncer se conforma con la frecuencia de replicación.
La dependencia del riesgo por célula con $N_{gen}$ es lineal, con una pendiente $4\times10^{-14}$, cuatro órdenes menor que la probabilidad de mutación $p_{mut}$
estimada en el EELP. Asumiendo, sin ninguna base desde luego, que células potencialmente cancerosas son generadas con frecuencia $p_{mut}$, resultaría 
que solo una de 10 000 continuaría su ruta hacia la formación del tumor. El resto sería eliminado por el sistema inmunológico.

\section{Probabilidades de impacto de las radiaciones sobre las células madres del pulmón}

A partir de los datos de la actividad del radón en la atmósfera tomados de la referencia \cite{Radon}, podemos ofrecer un 
estimado simple de la probabilidad de impacto de las part\'iculas $\alpha$
con las células madres del pulmón, provenientes de la desintegración del radón que respiramos. 

El tiempo de vida medio del isótopo $^{222}Rn$ es de 3.8 días, el cual es $100\%$ abundante naturalmente, como se dijo con anterioridad. 
El ritmo de flujo de aire en los pulmones, por otro lado, es de 4 litros por minuto. Teniendo en cuenta que la capacidad 
pulmonar es de aproximadamente de 3.5 a 4 litros, prácticamente todo el aire de los pulmones es renovado en cada minuto. Comparando el ritmo de respiración 
con el tiempo de vida medio del radón, en buena aproximación, la actividad de este isótopo en los pulmones es semejante a la encontrada en el aire circundante:
$40 Bq/m^{3}$ al aire libre y $150 Bq/m^{3}$ dentro de las casas.

\begin{figure}
\begin{center}
\includegraphics[width=9cm]{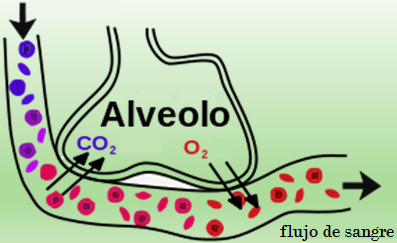}
\caption{Alveolo pulmonar y microcapilar sangu\'ineo. La membrana del alveolo es unicelular, de ancho menor que una micra, para permitir
el intercambio gaseoso. El di\'ametro del microcapilar es de $100\mu m$ aproximadamente.} \label{alveolo}
\end{center}
\end{figure}

Por otro lado, en el pulm\'on existen estructuras bronquiales llamadas alveolos, donde se almacena el aire inspirado y se 
realiza el intercambio gaseoso con la sangre, mediante la difusi\'on (\fig{alveolo}). La pared celular de los alveolos
es una capa unicelular, t\'ipicamente de menos de $1\mu m$, para que se de la difusi\'on. Los capilares que rodean a los alveolos tienen di\'ametros
de $100\mu m$ aproximadamente.

Una pregunta interesante es cu\'an profundo logran penetrar las part\'iculas $\alpha$ procedentes de la desintegración del rad\'on en nuestro organismo. 
Un estimado simple consiste en tomar datos de la potencia de frenado ($p_f$) de part\'iculas $\alpha$ en el agua presentes en \cite{stop}, para calcular 
su longitud
de penetraci\'on en el agua. Esta longitud no debe variar mucho en el caso de la sangre u otros tejidos, ya que su densidad es muy parecida a la del agua. La potencia
de frenado depende de la energ\'ia inicial que tengan las part\'iculas, $5.6 MeV$ en la desintegraci\'on del $^{222}Rn$. La potencia de frenado en este caso 
es de $p_f=4.32\times 10^{-3}g/cm^2$. El c\'alculo queda de la siguiente manera:

\begin{equation}
 l= \frac{p_f}{\rho} \approx 43.2 \mu m
\end{equation}

\noindent
donde la densidad del agua se tom\'o de $\rho=1g/cm^3$.
Comparando el valor de la longitud de penetraci\'on con las dimensiones de la pared celular que encuentran las part\'iculas $\alpha$ al 
introducirse en los alveolos,
podemos concluir que o bien se queda en el alveolo, o pasa a la sangre, en cualquier caso permanece en el pulm\'on.

Estableciendo que cada part\'icula $\alpha$ resultante de la desintegración del rad\'on impacta en alguna de sus c\'elulas y, teniendo en cuenta
que la proporci\'on de c\'elulas madres que hay en los pulmones es del $0.4\%$ (tomadas de la \cite{Tomasetti}), podemos calcular la frecuencia con que 
se impacta alguna c\'elula madre, denotada $F_{imp}$:

\begin{equation}
 F_{imp} = 0.004AV_p \sim 10^{-3}impactos/s
\end{equation}

\noindent
donde $A$ y $V$ son respectivamente la actividad del rad\'on en los pulmones (tomada de $100Bq$) y su volumen($4lt$). 
Tambi\'en podemos calcular la frecuencia por c\'elula en funci\'on de la cantidad de generaciones de las c\'elulas madres, para ello necesitamos 
la cantidad total de c\'elulas madres ($N_{celM}=1.2\times10^9$) y
el tiempo que tarda en dividirse cada una: $\tau=14.3$ a\~nos.

\begin{equation}
 f_{imp} =  F_{imp} \tau/N_{celM} \approx 4\times10^{-4}   impactos/gen \label{f1}
\end{equation}

Anteriormente hab\'iamos estimado que la probabilidad de ocurrencia de una mutaci\'on en una base del ADN de una c\'elula era de $10^{-9}-10^{-10}$ por
generaci\'on,
por lo que la probabilidad de que ocurra en un segmento de longitud $l$ ser\'a: $P_{rob}\sim10^{-9}l$. Nos interesan las mutaciones en los sectores 
codificantes del ADN, espec\'ificamente de los genes p53 y Rb, que son los que se sospecha que juegan un papel importante en la g\'enesis del c\'ancer.
En la Ref \cite{p53} aparecen datos de la longitud de estos genes, $l=10Kpb$ y $l=100Kpb$ ser\'ian estimados de los tamaños de la zona codificante.
La frecuencia de ocurrencia de mutaciones en estos genes ser\'ia de:

\begin{equation}
 f_{mut} =  10^{-9}\times100Kpb \sim 10^{-4}- 10^{-5}  mutaciones/gen \label{f2}
\end{equation}

\noindent
A pesar de la simplicidad del c\'alculo, las frecuencias de (\ref{f1}) y (\ref{f2}) son parecidas, lo que pudiera indicar que 
una de las causas fundamentales de la ocurrencia de mutaciones en las células madres pulmonares sea la desintegración del radón 
presente en el aire que respiramos.


\chapter{Conclusiones y Recomendaciones}\label{Cap4}

\section{Conclusiones}

A lo largo de este trabajo se han expuesto varias ideas relacionadas con las mutaciones en linajes celulares de bacterias y tejidos humanos. 
Se ha hecho una extensa revisi\'on de literatura sobre diversos t\'opicos, pero no en todos hemos llegado a resultados propios. Los principales
resultados de la tesis son los siguientes:

\begin{enumerate}

 \item Mostramos, a partir de datos del EELP, que las mutaciones en bacterias se pueden des\-cri\-bir por trayectorias de Levy, que se componen por mutaciones
 puntuales y reordenamientos no locales del ADN. La tasa de ambos eventos se estim\'o en $p_{mut}\approx10^{-10}$ mutaciones/generaci\'on, mientras que el par\'ametro 
 de Levy se estim\'o como $\nu \approx 3/2$.
 
 Pensamos que el resultado anterior tiene un car\'acter general, es decir, el modelo pudiera ser v\'alido para las mutaciones celulares, aunque los par\'ametros
 variar\'an en dependencia del sistema en cuesti\'on y de las condiciones a las que se encuentra expuesto. 
 
 Con estos resultados se ha elaborado un art\'iculo \cite{DarioyAugusto}.
 
 \item Se dise\~n\'o y puso a punto un algoritmo para mezclar las mutaciones con la selecci\'on natural en condiciones muy similares a las del EELP. Para ello se
 introdujo el  {\it fitness} de los clones. La simulaci\'on mostr\'o que la tasa de mutaciones beneficiosas que no se fijan es muy alta ($94.6\%$), por lo que el 
 estimado $p_{mut}\approx10^{-10}$ mutaciones/generaci\'on constituye una cuota inferior y el valor real en el experimento pudiera ser varias veces mayor.
 
 \item Se determin\'o la distribuci\'on del n\'umero de mutaciones y del {\it fitness} dentro de la poblaci\'on de bacterias, como funci\'on del n\'umero de 
 generaciones.  A partir de ellas se observan resultados interesantes que describen fen\'omenos del EELP, como la interferencia clonal y la epistasis.
 Estos resultados est\'an pendientes de elaborarse para publicaci\'on. 
 
 \item A partir de la distribuci\'on del {\it fitness} obtenemos una descripci\'on de la variedad genot\'ipica resultante de la competitividad de los clones.  
 Dicha variedad pudiera ser comprobada analizando las secuencias de ADN de las poblaciones de bacterias en el EELP.
 
\end{enumerate}

\section{Recomendaciones}

Las recomendaciones son, b\'asicamente, continuar las l\'ineas que han quedado pendientes:

\begin{enumerate}

 \item En el trabajo \cite{Long_Tail}, por ejemplo, se relacion\'o el genotipo mutante en el EELP con los reordenamientos no locales del ADN bacterial. 
 Esto no necesariamente es as\'i. En la referencia \cite{antibiotico}, los autores se\~nalan 
 a una mutaci\'on puntual en el gen Mut de {\it E. Coli} como la responsable de la 
 aparici\'on del genotipo mutante en algunas de las poblaciones. Lo indiscutible es que es un mutaci\'on beneficiosa que, al dar lugar a su vez a un incremento
 en la tasa de mutaciones, aumenta las posibilidades de hallar variantes con mayor {\it fitness}. Una posible l\'inea de trabajo consiste en correr 
 el algoritmo de competencia clonal en presencia de un mutante.
 
 \item De forma similar, en el trabajo \cite{Levy_Cancer} se relacion\'o la aparici\'on de c\'ancer con reordenamientos no locales del ADN. Esto tampoco 
 es as\'i necesariamente. La idea m\'as aceptada es que su g\'enesis es m\'ultiple \cite{Frank}, diversos cambios gen\'eticos o epigen\'eticos pueden 
 originarlo. Incluso existen opiniones de que un cambio, puntual o no, en uno de los genes p53 y Rb \cite{UnCambio} son capaces de desencadenar el tumor.
 Estos cambios pudieran ser debido a factores externos o intr\'insecos, pero, continuando el s\'imil con el genotipo mutante de {\it E. Coli}, su evoluci\'on
 depende de que la mutaci\'on le de cierta ventaja a la c\'elula en condiciones de homeostasis y vigilancia inmunol\'ogica del tejido. El problema es 
 interesante, pero en estos momentos no sabemos c\'omo modelarlo.
 
 \item Otra l\'inea de inter\'es es seguir investigando formas de medir los cambios que se producen en el ADN, para lograr as\'i, una comprensi\'on m\'as
 profunda de las mutaciones. En particular, ser\'ia interesante incorporar en esta descripci\'on, el control de la 
 expresi\'on gen\'etica; ya que, como hemos dicho antes, juega un papel fundamental en la g\'enesis del c\'ancer. 
 
 \item Hicimos una amplia revisi\'on del riesgo de c\'ancer y la dosis de radiaci\'on recibida por los distintos tejidos del cuerpo humano. Los datos
 m\'as elaborados que hallamos, se refieren a la poblaci\'on de los Estados Unidos \cite{Rep160} e indican al pulm\'on como el tejido sobre el cual
 la radiaci\'on tiene mayor impacto. Incluso existe un reporte detallado correlacionando especialmente el riesgo de c\'ancer de pulm\'on, en 
 diferentes estados, con las concentraciones naturales de rad\'on \cite{Radon}. Los efectos, sin embargo, son peque\~nos comparados con otros agentes 
 carcinog\'enicos. La radiaci\'on puede, por ejemplo, duplicar el riesgo de c\'ancer de pulm\'on, pero el h\'abito de fumar lo incrementa en 20 veces.
 De todas formas, el problema tiene gran importancia para la salud y se deber\'ia continuar.

 \item Finalmente, hicimos tambi\'en una revisi\'on de la exposici\'on de la poblaci\'on a distintas sustancias qu\'imicas, hallando datos muy meticulosos del 
 Departamento de Salud de los Estados Unidos \cite{Quimica}. A diferencia de las radiaciones, no existe aqu\'i un concepto an\'alogo al de dosis equivalente, 
 donde el efecto de las distintas sustancias pudiera compararse en una misma magnitud o par\'ametro. Debido al important\'isimo efecto t\'oxico y carcinog\'enico que tienen muchas 
 sustancias, se deber\'ia hacer un esfuerzo en sistematizar y caracterizar sus efectos. Es un terreno relativamente nuevo para la investigaci\'on y con gran impacto 
 para la salud. 
 
\end{enumerate}


\appendix
\chapter*{Anexos}\label{apA1}
\addcontentsline{toc}{chapter}{Anexos}

\chapter{Datos de riesgo de c\'ancer en tejidos} \label{apA}

\begin{table}[!ht]
\label{tab_Sol}
\begin{center}
\begin{tabular}{|l|c|c|c|c|}
  \hline
  \textbf{Tipo de c\'ancer} &  \textbf{Riesgo} & \textbf{Cantidad de} & \textbf{Cantidad de}&\textbf{Cantidad de}\\ 
  &\textbf{por vida}&\textbf{c\'elulas$^*$ del}&\textbf{c\'elulas madres$^*$}&\textbf{divisiones de}\\
  &&\textbf{tejido donde}&\textbf{en el tejido}&\textbf{c\'elulas madres}\\
  &&\textbf{se origina}&&\textbf{por a\~no}\\ \hline
  leucemia mieloide aguda &  0.0041 & $3 \times 10^{12}$ & $1.35 \times 10^{8}$ &12\\ \hline
  carcinoma basocelular &  0.3 & $1.8 \times 10^{11}$ & $5.82 \times 10^{9}$ &7.6\\ \hline
  leucemia linf\'atica cr\'onica &  0.0052 & $3 \times 10^{12}$ & $1.35 \times 10^{8}$ &12\\ \hline
  c\'ancer colorrectal &  0.048 & $3 \times 10^{10}$ & $2 \times 10^{8}$ &73\\ \hline
  c\'ancer colorrectal con PAF &  1 & $3 \times 10^{10}$ & $2 \times 10^{8}$ &73\\ \hline
  c\'ancer colorrectal con&  0.5 & $3 \times 10^{10}$ & $2 \times 10^{8}$ &73\\ 
  s\'indrome de Lynch &&&&\\ \hline
  c\'ancer de duodeno &  0.0003 & $6.8 \times 10^{8}$ & $4 \times 10^{6}$ &24\\ \hline
  c\'ancer de duodeno con PAF&  0.035 & $6.8 \times 10^{8}$ & $4 \times 10^{6}$ &24\\ \hline
  c\'ancer de c\'elulas escamosas&  0.001938 & $3.24 \times 10^{9}$ & $8.64 \times 10^{5}$ &17.4\\ 
  del es\'ofago&&&&\\ \hline
  c\'ancer de ves\'icula  &  0.0028 & $1.6 \times 10^{8}$ & $1.6 \times 10^{6}$ &0.584\\ 
  no papilario&&&&\\ \hline
  glioblastoma &  0.00219 & $8.46 \times 10^{10}$ & $1.35 \times 10^{8}$ &0\\ \hline
  
\end{tabular}
\end{center}
\caption{Datos de riesgo de incidencia de c\'ancer en la vida y par\'ametros relacionados 
con las c\'elulas madres normales que son precursoras de esos c\'anceres \cite{Tomasetti}. $^*$ ``C\'elulas'' y ``c\'elulas madres'' se refiere solamente
a aquellas c\'elulas normales del mismo tipo que las c\'elulas cancerosas de ese tejido. Por ejemplo, para el c\'ancer colorrectal, las
c\'elulas y las c\'elulas madres se refieren a c\'elulas epiteliales, no a los estromas u otros tipos de c\'elulas de colon an\'omalo.}
\end{table}  
  
\begin{table}[!ht]
\label{tab_Sol1}
\begin{center}
\begin{tabular}{|l|c|c|c|c|}
  \hline

  carcinoma de c\'elulas &  0.0138 & $1.67 \times 10^{10}$ & $1.85 \times 10^{7}$ &21.5\\ 
  escamosas de la cabeza&&&&\\
  y el cuello&&&&\\ \hline
  carcinoma de c\'elulas &  0.07935 & $1.67 \times 10^{10}$ & $1.85 \times 10^{7}$ &21.5\\ 
  escamosas de la cabeza&&&&\\ 
  y el cuello con VPH-16&&&&\\ \hline
  hepatocarcinoma &  0.0071 & $2.41 \times 10^{11}$ & $3.01 \times 10^{9}$ &0.9125\\ \hline
  hepatocarcinoma &  0.071 & $2.41 \times 10^{11}$ & $3.01 \times 10^{9}$ &0.9125\\ 
  con hepatitis C&&&&\\ \hline
  adenocarcinoma de pulm\'on &  0.0045 & $4.34 \times 10^{11}$ & $1.22 \times 10^{9}$ &0.07\\ 
  (no fumadores)&&&&\\ \hline
  adenocarcinoma de pulm\'on &  0.081 & $4.34 \times 10^{11}$ & $1.22 \times 10^{9}$ &0.07\\ 
  (fumadores)&&&&\\ \hline
  meduloblastoma &  0.00011 & $8.5 \times 10^{10}$ & $1.36 \times 10^{8}$ &0\\ \hline
  melanoma &  0.0203 & $3.8 \times 10^{9}$ & $3.8 \times 10^{9}$ &2.48\\ \hline
  osteosarcoma &  0.00035 & $1.9 \times 10^{9}$ & $4.18 \times 10^{6}$ &0.067\\ \hline
  osteosarcoma de los brazos &  0.00004 & $3 \times 10^{8}$ & $6.5 \times 10^{5}$ &0.067\\ \hline
  osteosarcoma de la cabeza &  0.0000302 & $3.9 \times 10^{8}$ & $8.6 \times 10^{5}$ &0.067\\ \hline
  osteosarcoma de las piernas &  0.00022 & $7.2 \times 10^{8}$ & $1.59 \times 10^{6}$ &0.067\\ \hline
  osteosarcoma de la pelvis &  0.00003 & $2 \times 10^{8}$ & $4.5 \times 10^{5}$ &0.067\\ \hline
  c\'elulas germinales del ovario &  0.000411 & $1.1 \times 10^{7}$ & $1.1 \times 10^{7}$ &0\\ \hline
  adenocarcinoma de c\'elulas de &  0.013589 & $1.672 \times 10^{11}$ & $4.18 \times 10^{9}$ &1\\ 
  los conductos pancre\'aticos&&(\'acinos)&&\\ \hline
  carcinoma neuroendocrino pancre\'atico &  0.000194 & $2.95 \times 10^{9}$ (islote)& $7.4 \times 10^{7}$ &1\\ \hline
  adenocarcinoma del intestino delgado &  0.0007 & $1.7 \times 10^{10}$ & $1 \times 10^{8}$ &36\\ \hline
  c\'elulas germinales de los test\'iculos &  0.0037 & $2.16 \times 10^{10}$ & $7.2 \times 10^{6}$ &5.8\\ \hline
  carcinoma folicular/papilar de tiroides &  0.01026 & $ 10^{10}$ & $6.5 \times 10^{7}$ &0.087\\ \hline
  carcinoma de tiroides medular &  0.000324 & $ 10^{9}$ & $6.5 \times 10^{6}$ &0.087\\ \hline
\end{tabular}
\end{center}
\caption{Continuaci\'on de la Tabla A.1}
\end{table}



\chapter{Coeficientes de la dosis}\label{apB}
\section{Peso de la radiación}\label{apB1}

Para la descripción ...

\begin{table}[!ht]
\label{tab_Sol3}
\begin{center}
\begin{tabular}{|l c|}
  \hline
  \textbf{Tipo de radiaci\'on} &  \textbf{factor $w_R$} \\ \hline
  Fotones &$1$\\
  Electrones y muones &$1$\\
  Protones y piones cargados &$2$\\
  Partículas alfa ,Fragmentos de fisión, Iones pesados &$20$\\
  Neutrones &Una función continua de la energía del\\
  &neutrón: $w_R(E_n)$\\ \hline
  
\end{tabular}
\end{center}
\caption{ Factores de ponderación recomendados para la radiación.}
\end{table}  

\begin{figure}
 \begin{center}
\includegraphics[width=18cm]{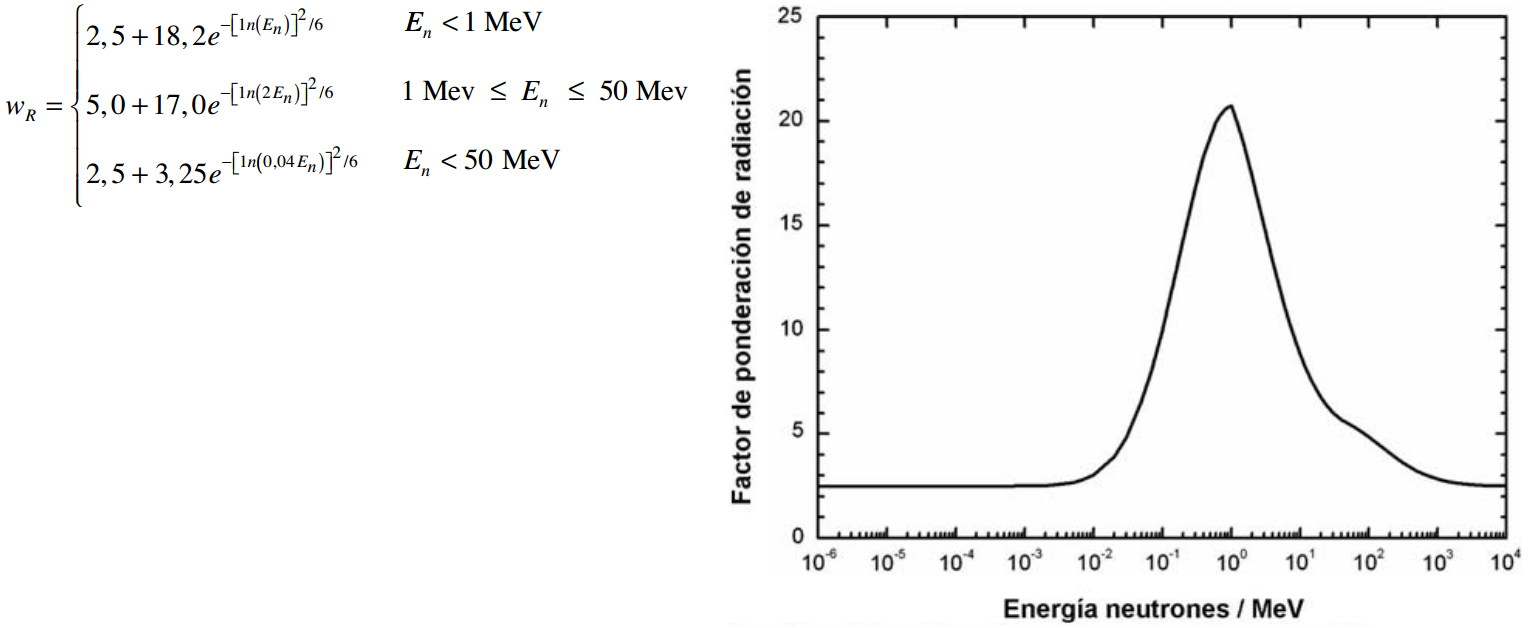}
\caption{ Factor de ponderación de los neutrones como funci\'on de su energía.} \label{Bp1}
\end{center} 
\end{figure}

El factor de ponderación de la radiación para los neutrones refleja su eficacia biológica relativa debida a la exposición externa. 
La eficacia biológica de neutrones incidentes en el cuerpo
humano depende fuertemente de la energía del neutrón (\fig{Bp1}).

\section{Sensibilidad del tejido}\label{apB2}

\begin{table}[!ht]
\label{tab_Sol4}
\begin{center}
\begin{tabular}{|l c c|}
  \hline
  \textbf{Tejido} &  \textbf{$w_T$} & \textbf{$\sum{w_T}$}\\ \hline
  Medula ósea, colon, pulmón, estómago, mama, resto de los tejidos $^*$ &$0.12$&$0.72$\\
  Gónadas &$0.08$ &$0.08$\\
  Vejiga, esófago, hígado, tiroides &$0.04$ & $0.16$\\
  Superficie del hueso, cerebro, glándulas salivales, piel &0.01 & 0.04\\ 
  & Total & 1.00\\ \hline
  
\end{tabular}
\end{center}
\caption{ Factores de ponderación recomendados para los tejidos. $^*$Resto de los Tejidos: Adrenales, región extra torácica, vesícula, corazón, 
riñones, nódulos linfáticos, músculo, mucosa oral, páncreas, próstata, intestino delgado, bazo, timo, útero/cérvix.
}
\end{table}



\begin{thebibliography}{99}
\addcontentsline{toc}{chapter}{Bibliografía.}
\markright{Bibliografía.}

\bibitem{BioMol} B. Alberts, A. Johnson, J. Lewis, M. Raff, K. Roberts, P. Walter, Molecular Biology of the Cell, (New York: Garland Science, 2002). 

\bibitem{20mil_Lenski} R.E. Lenski, Phenotypic and genomic evolution during a 20000 generation experiment with the bacterium E. Coli, in J. Janick, Ed., 
Plant Breeding Reviews, Vol. 24, Part 2 (2004), p. 225.

\bibitem{Adaptation_Lenski} Michael J. Wiser, Noah Ribeck, Richard E. Lenski, Long-Term Dynamics of Adaptation in Asexual Populations, Science 342 1364 (2013).

\bibitem{Fitness} H. Allen Orr, Fitness and its role in evolutionary genetics, Nat Rev Genet (2009) 10(8): 531-539.

\bibitem{Sitio_Lenski} Richard E. Lenski, Sumary data from the long-term evolution experiment, http://myxo.css.msu.edu/ecoli/summdata.html (2016).

\bibitem{A1} J.E. Barrick, R.E. Lenski, Cold Spring Harbor Symposia on Quantitative Biology, Vol. 54, 1 (2009).

\bibitem{A2} J.E. Barrick, D. Su Yu, S. Ho Yoon, H. Jeong, T. Kwang Oh, D. Schneider, R.E. Lenski, J.F. Kim, Genome evolution and adaptation in a
long-term experiment with Escherichia coli, Nature Vol. 461 (2009).

\bibitem{mBio} C. Raeside, J. Gaffé, D.E. Deatherage, O. Tenaillon, A.M. Briska, R.N. Ptashkin, S. Cruveiller, C. Médigue, R.E. Lenski, J.E. Barrick, 
D. Schneider, Large chromosomal rearrangements during a long-term evolution experiment with Escherichia coli, mBio (2014) 5(5): e01377-14. 

\bibitem{Tomasetti} C. Tomasetti, B. Vogelstein, Variation in cancer risk among tissues can be explained by the number of stem cell divisions, Science 
347 (2015), pp. 78-81.

\bibitem{w} Recomendaciones de la Comisión Internacional de Protección Radiológica,  editada por la Sociedad Española de Protección
Radiológica con la autorización de la International Commission on Radiological Protection (ICRP), ICRP n\'umero 103 (2007).

\bibitem{Rep160} Ionizing Radiation Exposure of the Population of the United States, Recommendations of the
National Council on Radiation Protection and Measurements, NCRP report no. 160,  http://NCRPpublications.org (2009).


\bibitem{Radon} Health Effects of Exposure to Radon, Committee on Health Risks of Exposure to Radon (BEIR VI), National Research Council, 
http://www.nap.edu/catalog/5499.html (1999).

\bibitem{Long_Tail} A. González, The long-tail distribution function of mutation in bacteria, Revista Cubana de Física, Vol. 32, 86 (2015).

\bibitem{Levy_Cancer} A. González, Levy model of cancer, arXiv:1507.08232.

\bibitem{Probability} V. S. Koroliuk, N. I. Portenko, A. V. Skorojod, A. F. Turbin, Handbook on probability theory and mathematical statistics, (Nauka, 
Moscow, 1978).

\bibitem{Landscape} A. L.  Goldberger, S. Havlin,  R.N. Mantegna,  M. E. Matsa, C. K. Peng, M.  Simons, H. E. Stanley, Long-range correlation properties
of coding and noncoding DNA sequences: GenBank analysis, Physical Review E, Vol. 51, número. 5, (1995).

\bibitem{Walks} C. K.  Peng, S. V.  Buldyrev, A. L.  Goldberger, F.  Sciortino, M.  Simons, H. E.  Stanley, Fractal landscape analysis of DNA walks, 
Physica A 191 (1992) pp. 25-29. 

\bibitem{Entropy} T. D. Schneider, Information and entropy of patterns in genetic switches, In G. J. Erickson and C. R. Smith, Eds., Maximum-Entropy, 
Bayesian Methods in Science and Engineering, Vol. 2, (Dordrecht, Kluwer Academic, 1988), pp. 147-154. 

\bibitem{Levy_Walks} Sergey V. Buldyrev, Generalized Levy-walk model for DNA nucleotide sequences, Physical Review E, Vol. 47, número. 6, (1996).

\bibitem{Levy_Flights} Eds. M. F. Shlesinger, G. Zaslavsky, U. Frish, Levy flights and related phenomena in Physics, Lecture Notes in Physics, Springer, 
Vol. 450, (Berlin, 1995).

\bibitem{Computacion} Chang-Yong Lee, Xin Yao, IEEE Trans. Evol. Comp., Vol. 8, No. 1, pp 1, (2004).

\bibitem{Notes_Fitness} S. Nuismer, Lecture Notes for Evolutionary Ecology 548. Lecture $\#2$: Fitness, Selection, and Adaptation, (University of Idaho 2011).

\bibitem{antibiotico} Oliver A., Canton R., Campo P., Baquero F., Blazquez J., High frequency of hypermutable Pseudomonas aeruginosa in
cystic fibrosis lung infection. Science 288 (2000) p1251 - 1254.


\bibitem{Frank} Steven A. Frank, Dynamics of Cancer: Incidence, Inheritance, and Evolution, Princeton series in evolutionary biology (U.K. 2007),
 http://creativecommons.org/licenses/by-nc/2.5/
 
\bibitem{stop} ASTAR: Stopping power and range tables for alpha particules, http://physics.nist.gov/cgi-bin/star/ap-table.pl
 
\bibitem{p53} Genome Browser Gateway University of California Santa Cruz, http://genome.uscs.edu 

\bibitem{DarioyAugusto} Dario Le\'on, Augusto González, Mutations as Levy flights, arXiv: 1605.09697

\bibitem{UnCambio} L.L. Mays Hoopes, Aging and cell division, Nature Education 3 (2010) p55.

\bibitem{Quimica} Fourth National Report on Human Exposure to Environmental Chemicals, Department of Health and Human Services Centers of Disease 
Control and Prevention, http://www.cdc.gov/exposurereport (2009).

\end{thebibliography}

\end{document}